\documentclass[conference]{IEEEtran}

\usepackage{cite}
\usepackage{amsmath,amssymb,amsfonts}
\usepackage{algorithmic}
\usepackage{graphicx}
\usepackage{textcomp}
\usepackage{xcolor}
\usepackage{fancyhdr}
\usepackage[hyphens]{url}

\usepackage{multicol}
\usepackage[position=top]{subfig}
\usepackage{algorithm}
\usepackage{colortbl}
\usepackage{color,soul}

\def\BibTeX{{\rm B\kern-.05em{\sc i\kern-.025em b}\kern-.08em
    T\kern-.1667em\lower.7ex\hbox{E}\kern-.125emX}}

\begin{document}

\title{CREW: Computation Reuse and Efficient Weight Storage for Hardware-accelerated MLPs and RNNs}
\author{Marc Riera, Jose-Maria~Arnau, Antonio Gonzalez}
\maketitle


\begin{abstract}
Deep Neural Networks (DNNs) have achieved tremendous success for cognitive applications. The core operation in a DNN is the dot product between quantized inputs and weights. Prior works exploit the weight/input repetition that arises due to quantization to avoid redundant computations in Convolutional Neural Networks (CNNs). However, in this paper we show that their effectiveness is severely limited when applied to Fully-Connected (FC) layers, which are commonly used in state-of-the-art DNNs, as it is the case of modern Recurrent Neural Networks (RNNs) and Transformer models.

To improve energy-efficiency of FC computation we present CREW, a hardware accelerator that implements Computation Reuse and an Efficient Weight Storage mechanism to exploit the large number of repeated weights in FC layers. CREW first performs the multiplications of the unique weights by their respective inputs and stores the results in an on-chip buffer. The storage requirements are modest due to the small number of unique weights and the relatively small size of the input compared to convolutional layers. Next, CREW computes each output by fetching and adding its required products. To this end, each weight is replaced offline by an index in the buffer of unique products. Indices are typically smaller than the quantized weights, since the number of unique weights for each input tends to be much lower than the range of quantized weights, which reduces storage and memory bandwidth requirements.

Overall, CREW greatly reduces the number of multiplications and provides significant savings in model memory footprint and memory bandwidth usage. We evaluate CREW on a diverse set of modern DNNs. On average, CREW provides $2.61x$ speedup and $2.42x$ energy savings over a TPU-like accelerator. Compared to UCNN, a state-of-art computation reuse technique, CREW achieves $2.10x$ speedup and $2.08x$ energy savings on average.
\end{abstract}

\section{Introduction}\label{s:intro}
Deep Neural Networks (DNNs) represent the state-of-the-art solution to a broad range of applications such as machine translation~\cite{gnmt} and speech recognition~\cite{deepspeech2}. The complexity of the DNN models continues to grow along with its computational cost and memory requirements, making it difficult to support them on conventional computer systems. Accelerators adopting a systolic array architecture such as TPU~\cite{TPU} from Google, have proven to be efficient hardware implementations to perform DNN inference~\cite{Eyeriss, SCNN, FlexFlow}. The systolic array architecture~\cite{systolic, unified_systolic_ann}, which was designed for massive parallelization and data reuse, is especially effective for Convolutional Neural Networks (CNNs) since the weights of a layer are shared across a large number of sliding windows and can be reused multiple times. In addition, there has been a plethora of recent proposals to further optimize CNN inference~\cite{Morph, UCNN, SnaPEA}. However, CNNs are just a subset of DNNs with very specific characteristics. Therefore, techniques targeting CNNs do not necessarily achieve similar benefits for other DNN architectures. Our proposal is motivated by the fact that state-of-the-art models for sequence-to-sequence problems are either Recurrent Neural Networks (RNNs)~\cite{gnmt, deepspeech2} or very deep Multi-Layer Perceptrons (MLPs) such as the Transformer~\cite{attention}, which are both composed of fully-connected (FC) layers. FC layers exhibit different characteristics with respect to CNNs: weights are not reused by different neurons and the compute to memory access ratio is significantly smaller, i.e., FC layers are more memory intensive.

\begin{figure}[t!]
\centering
\includegraphics[width=0.90\columnwidth]{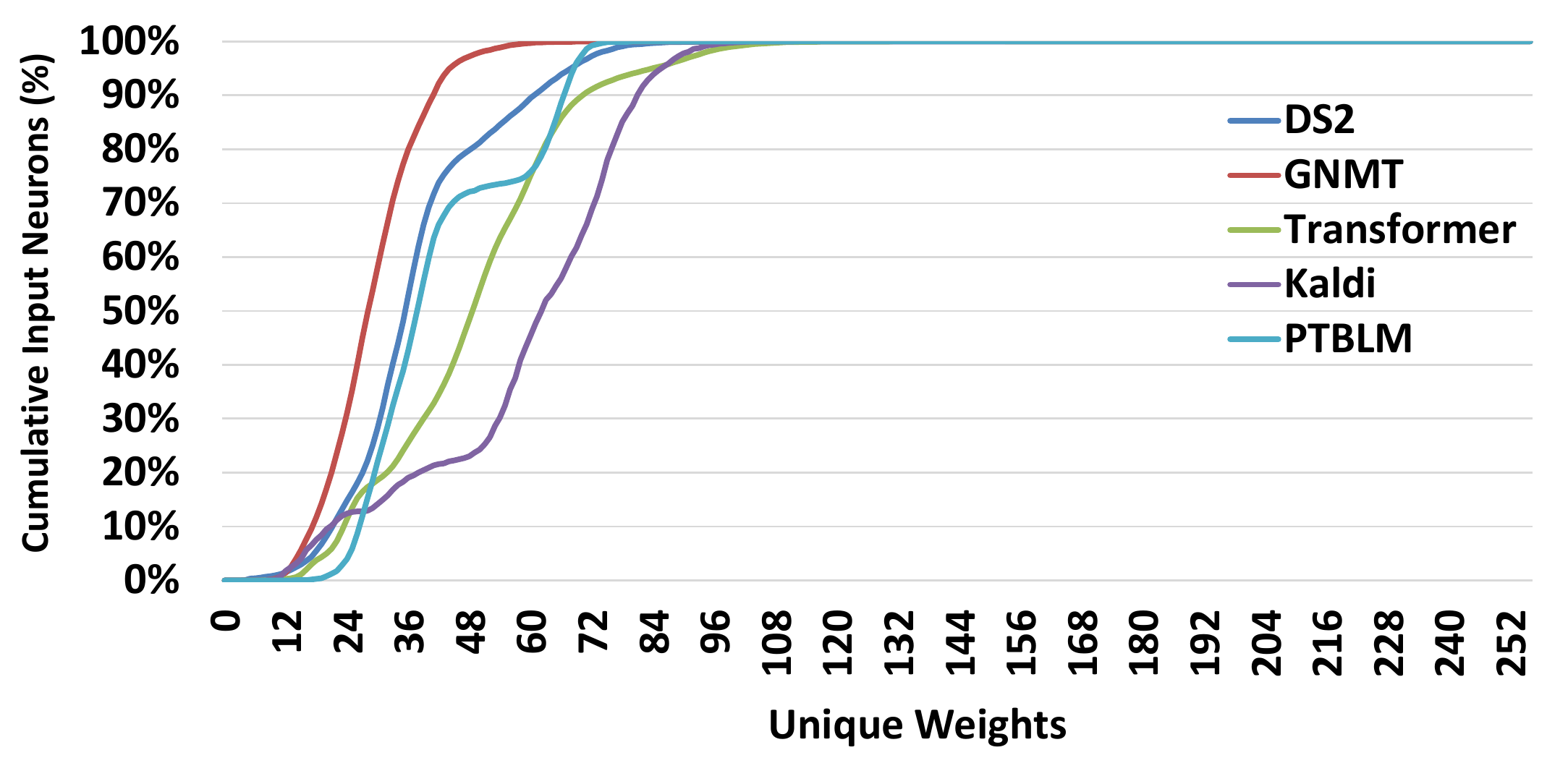}
\caption{Cumulative distribution of the unique weights per input neuron from all the FC layers of DeepSpeech2~\cite{deepspeech2}, GNMT~\cite{gnmt}, Transformer~\cite{attention}, Kaldi~\cite{Kaldi} and PTBLM~\cite{PTBLM}.}
\vskip -0.25in
\label{f:all_uw_accu}
\end{figure}

\begin{figure*}[ht!]
    \centering
    \subfloat[Standard DNN FC Layer]{
        \includegraphics[width=0.85\columnwidth]{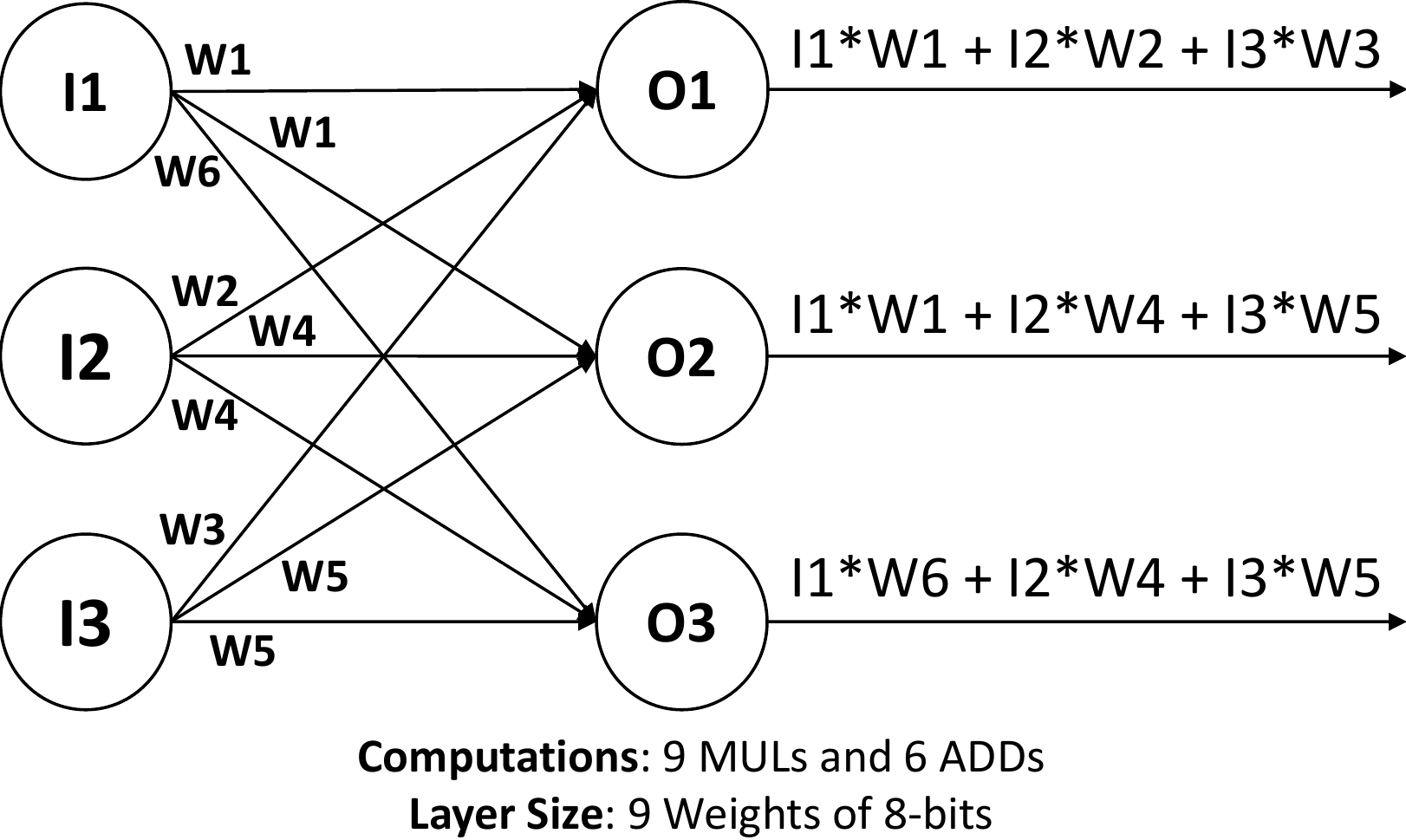}
        \label{f:dnn_fc}
    }
    \hspace{1.5cm}
    \subfloat[CREW FC Layer]{
        \includegraphics[width=0.75\columnwidth]{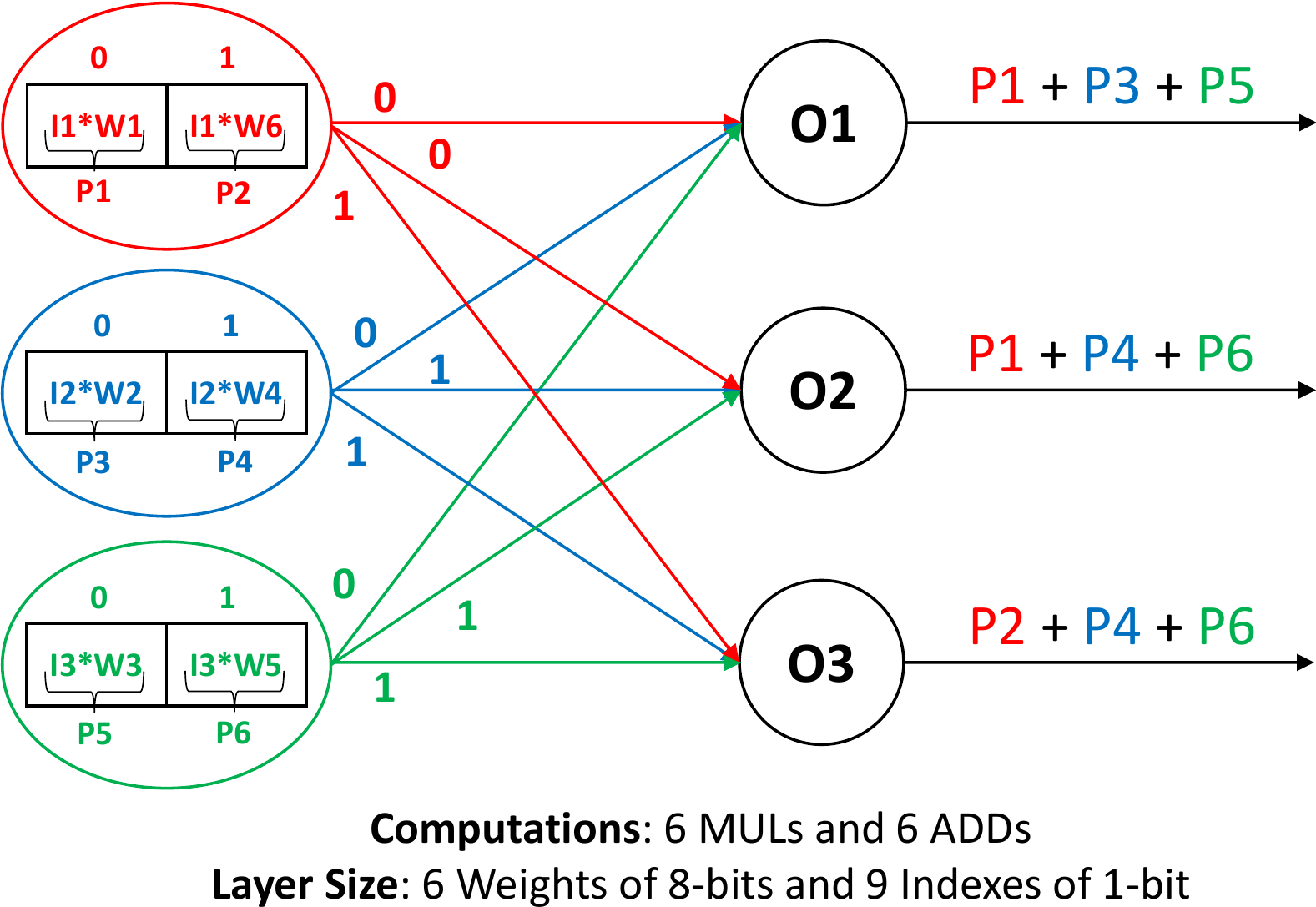}
        \label{f:crew_fc}
    }
    \caption{Standard (a) vs CREW (b) DNN inference for a fully-connected (FC) layer. For each input, CREW only performs the multiplications with unique weights, avoiding redundant computations and memory fetches. Furthermore, it replaces weights by indexes in the buffer of unique multiplications. In this example there are only two unique weights per input, so the 8-bit quantized weights can be replaced by 1-bit indexes.}
    \vskip -0.25in
    \label{f:fc_comp}
\end{figure*}

FC layer computation mainly consists of dot products between a vector of inputs and a matrix of weights. The inputs are generated dynamically while the weights are static and learned during the training phase. Current DNN model sizes are in the hundreds of megabytes (MB) and require billions of floating-point (FP) computations to perform inference. Linear quantization~\cite{minerva, TPU} is a highly popular technique used to compress the weights of the model and reduce the complexity of the computations. Quantization maps a continuous set of values to a discrete set, which as a side effect favors the appearance of repeated weights. Our work is based on the observation that, on average, more than 80\% of the inputs among different FC layers of a set of DNNs are multiplied by less than $64$ unique weights. Figure~\ref{f:all_uw_accu} shows the cumulative distribution of unique weights per input of DeepSpeech2 (DS2)~\cite{deepspeech2}, GNMT~\cite{gnmt}, Transformer~\cite{attention}, Kaldi~\cite{Kaldi} and PTBLM~\cite{PTBLM}, which are popular models for speech recognition, machine translation, and language modeling. The average number of unique weights per input is $44$ when applying an 8-bit quantization, and can never be higher than 256.

In this paper, we show how to efficiently exploit the spatial locality of repeated weights on FC layers. We first propose a mechanism (CREW) that dynamically computes and stores the partial products of each input by their associated unique weights in a given layer. This first step removes redundant multiplications due to repeated weights and their associated memory reads. Similar to sparse architectures~\cite{EIE, SCNN, CambriconX} and their pruning mechanisms~\cite{NearZeroPruning, Scalpel}, the weight repetition patterns are irregular, making it very challenging to achieve net benefits due to the sparse accesses and extra metadata that are required. The partial products stored by this mechanism have to be indexed similar to sparse algorithms. Since the unique weights are known statically, the indexation table can be generated offline. The second step of our mechanism produces the final outputs of a FC layer by accessing the partial products with the associated indices and adding all the required values for each output. The main advantage of our mechanism is that the size of the indices depends on the number of unique weights of each input. As previously described, the number of unique weights is typically lower than $64$ so indices will be typically smaller than $7$ bits. In consequence, the final size of the model can be further compressed and the memory bandwidth is further reduced. Figure~\ref{f:fc_comp} shows an example of a small FC layer computing the standard dot product (Figure~\ref{f:dnn_fc}) and our CREW mechanism (Figure~\ref{f:crew_fc}), saving 33\% of multiplications and 20\% of storage.

Then, we present CREW, a novel accelerator that implements the above computation reuse and efficient weight storage scheme for FC layers. CREW is implemented on top of a TPU-like architecture, but it includes an enhanced weight stationary dataflow. The extra hardware required for our technique is modest since most of the components are already available in the baseline TPU-like architecture. CREW only requires small local buffers to store blocks of indices and some shared buffers to store partial results. The extra memory represents a small increase in the on-chip storage of the accelerator. Our experimental results show that the overheads are minimal compared to the savings in memory fetches and multiplications. Although our scheme has some similarity with UCNN~\cite{UCNN}, a recently proposed mechanism for computation reuse in CNNs, our work is different in several manners. First, we observe the low number of unique weights per input neuron of different FC layers instead of per output neuron. Second, we exploit this observation to reduce the size of the indexation. UCNN focuses on convolutional layers and shows modest speedups for FC layers compared to CREW. Our scheme not only avoids most of the multiplications, it also provides a significant reduction in storage and memory accesses due to the smaller indexes compared to the original weights. To summarize, this paper focuses on energy-efficient FC layer inference. The main contributions are:

\begin{itemize}
\item We analyze the unique weights per input in FC layers for several state-of-the-art MLPs and RNNs, including a Transformer. We observe that, on average, each input is multiplied by $44$ unique weights.

\item We propose a novel computation reuse mechanism to exploit weight repetition on FC layers. This technique reduces the number of multiplications and memory accesses by 98\% and 40\% respectively on average. Furthermore, it replaces weights by indices in the buffer of partial results, resulting in 25\% reduction of the model size.

\item We present CREW, a hardware accelerator that implements our computation reuse scheme. CREW improves performance by $2.61x$ and reduces energy consumption by $2.42x$ on average over a TPU-like accelerator. Compared to UCNN~\cite{UCNN}, CREW achieves $2.10x$ speedup and $2.08x$ energy savings on average.
\end{itemize}

The rest of the paper is organized as follows. Section~\ref{s:background} provides background information on modern DNNs. Section~\ref{s:WR} reviews prior computation reuse mechanisms and motives our own proposal by presenting the analysis of unique weights per input. Section~\ref{s:partial_reuse} describes in detail our reuse scheme. Section~\ref{s:accelerator} describes the hardware implementation of our computation reuse and model compression scheme. Section~\ref{s:methodology} presents the evaluation methodology and Section~\ref{s:results} discusses the experimental results. Section~\ref{s:related_work} reviews some related work and, finally, Section~\ref{s:conclusions} sums up the main conclusions.

\section{Background}\label{s:background}

\subsection{Modern DNNs}\label{s:DNN}
Deep Neural Networks (DNNs) can be classified in three main categories. First, Multi-Layer Perceptrons (MLP)~\cite{SpeechMLP, MNT_MLP} consist of multiple Fully-Connected (FC) layers in which every input neuron is connected, via synapses with particular weights, to every output neuron. Second, Convolutional Neural Networks (CNN) are composed of multiple convolutional layers to extract features, usually followed by one or several FC layers to perform the final classification. CNNs have proved to be particularly efficient for image and video processing~\cite{AlexNetV1, ResNet, DenseNet, 3D_CONV}. Finally, Recurrent Neural Networks (RNN) consist of multiple layers of cells with feedback connections, stacked on top of each other. RNN cells store information from past executions to improve the accuracy of future predictions. The most popular RNN architectures are the Long-Short Term Memory (LSTM)~\cite{LSTM} and the Gated Recurrent Unit (GRU)~\cite{GRU}. In both cases, the cell consists of multiple single-layer FC networks commonly referred as gates.

RNNs and very deep MLPs such as the Transformer model have become the state-of-art solution for sequence processing problems such as machine translation and speech recognition~\cite{gnmt, deepspeech2, attention}. The Transformer~\cite{attention} model has recently received special attention from the machine learning community for being extremely efficient in terms of both accuracy and performance. Transformers use attention mechanisms~\cite{bahdanau_attention, luong_attention} to gather information about the relevant context of a given input (i.e., a word of a sentence), and then encode that context in a vector. The attention mechanism allows to grab context information from distant parts of an input sequence to help understand its meaning, and it is implemented in the form of multiple feed-forward FC layers.

Most recent proposals focus on optimizing CNN inference and its convolutional layers. Although CNNs are the most efficient solution for image processing applications, their unique characteristics make it hard to exploit the same techniques on other DNNs. In addition, state-of-art accelerators implementing systolic array architectures, such as TPU~\cite{TPU}, have proven to be efficient hardware implementations to perform CNN inference since the weights of a convolutional filter are reused multiple times. However, the FC layers do not have the same reuse properties and may cause the accelerator to be highly underutilized, especially for small batch sizes that are common in inference. In this paper we focus on optimizing the performance of hardware accelerators for FC layers inference in modern MLPs, RNNs and Transformers.

\subsection{Fully-Connected Layers}\label{s:FC}
The main performance and energy bottleneck of MLPs and RNNs are the FC layers. In an FC layer, each output neuron performs a dot product operation between all the inputs of the layer and their corresponding weights. Specifically, the output of neuron $j$ is computed according to the following equation:

\begin{equation}
out(j) = (\sum_{i=0}^{N-1} w_{ij} * in(i)) + b_j
\end{equation}

where $in(i)$ represents the input vector, $w_{ij}$ is the weight of neuron $j$ for input $i$, and $b_j$ represents the bias of neuron $j$. An FC layer with $N$ inputs and $M$ neurons contains $N \times M$ weights, as each neuron has its own set of weights, and requires $2 \times N \times M$ operations. FC layers employed in real applications consist of thousands of neurons and they typically account for most of the computations and memory bandwidth usage of MLPs and RNNs. FC layers may also take most of the storage requirements and memory bandwidth usage in CNNs that tend to include a few FC layers at the end of the model to perform the final classification.

\section{Unique Weights in FC Layers}\label{s:WR}
FC layers have increased their size over time from thousands to millions of parameters to provide better accuracy on complex applications. However, the number of unique weights in each neuron is decreasing. Prior works observed the same effect in the filters of the CNNs~\cite{UCNN}. This reduction in unique weights is due to the successful methods to compress the DNN model size such as pruning~\cite{NearZeroPruning, Scalpel} and quantization~\cite{dorefa, lqnets}. Linear quantization~\cite{original_qt} is a highly popular technique to map a continuous set of values to a discrete set with a negligible impact in accuracy. One of the side effects of quantization is that it significantly increases the number of repeated weights. In this paper, we apply uniformly distributed linear quantization to the weights of the FC layers. Then, we analyze the number of unique weights per input neuron.

Typically, DNNs are quantized to 8 bits per weight without any impact in accuracy~\cite{TPU}. We refer to the number of unique weights in a layer as $UW$, and hence with 8-bit weights $UW \leq 2^8 = 256$. Weight repetition in a fully-connected layer is guaranteed as long as $UW < N * M$, being $N$ and $M$ the number of input and output neurons of the FC layer respectively. This condition is commonly met in modern DNNs. For instance, all the FC layers of the Transformer~\cite{attention} have more than one million weights.

Weight repetition can be exploited in two different ways: Factorization and Memoization. UCNN~\cite{UCNN}, a state-of-art accelerator, implements a computation reuse technique by exploiting the factorization of repeated weights. UCNN focuses on optimizing the convolutional layers by grouping and adding the inputs that belong to the same unique weights in a given convolutional window and filter, performing each unique weight multiplication just once per factorization group. UCNN implements FC layers as convolutions, where each output neuron is treated as a $1 \times 1 \times N$ convolutional filter, i.e. the number of channels is the number of inputs ($N$) in the FC layer. Due to the factorization, inputs are spread out irregularly within each filter and, hence, an indirection table is required to fetch the inputs for each unique weight. Since there are $N$ inputs, the size of each index in the indirection table is $log_2 N$. Note that for FC layers in modern DNNs, $log_2 N$ may be larger than 8 bits. To sum up, UCNN reduces the number of computations in an FC layer. However, it requires $N \times M$ indexes of size $log_2 N$ bits, which may result in a model larger than the original one. Therefore, its indexing overheads to apply factorization on FC layers are not negligible and hinder the benefits of the computation reuse. Not surprisingly, we obtained modest speedups and energy savings when applying UCNN on FC layers as we show in Section~\ref{s:results}.

\begin{figure}[t!]
\centering
\includegraphics[width=1\columnwidth]{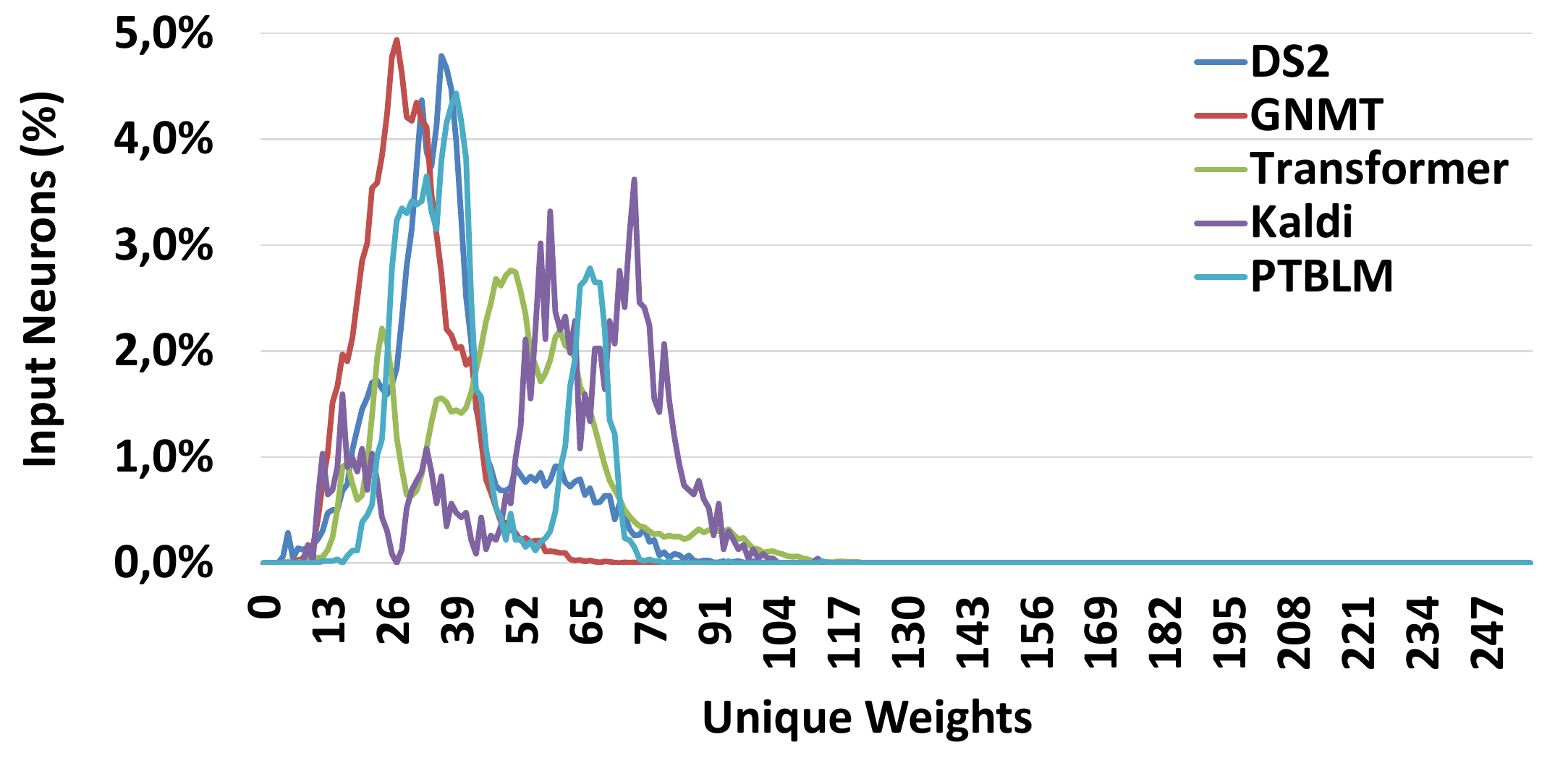}
\caption{Histograms of the unique weights per input neuron from all the FC layers of DS2, GNMT, Transformer, Kaldi and PTBLM.}
\vskip -0.10in
\label{f:all_uw_hist}
\end{figure}

On the other hand, we observe that the number of unique weights per input neuron is relatively low. Therefore, we can memoize partial products between inputs and unique weights to largely reduce the number of multiplications. The number of indexes to the partial results required is still $N * M$, but the size of each index will depend on the number of unique weights of its related input neuron. Therefore, if the condition in Equation~\ref{eq:mem_cond} is fulfilled, where $q$ is the number of bits used to quantize and $UW_i$ is the number of unique weights of a given input neuron $i$, the size of the model along with the associated memory accesses can be reduced.

\begin{equation}
\label{eq:mem_cond}
UW_i \leq 2^{q-1}, \forall i \in \{1,\ldots,N\}
\end{equation}

CREW, our proposed computation reuse solution for FC layers, is based on the key observation that, on average, more than 80\% of the input neurons among different FC layers of a representative set of DNNs are multiplied by less than $64$ unique weights as shown in Figure~\ref{f:all_uw_accu}. Figure~\ref{f:all_uw_hist} provides the histogram of unique weights per input neuron for each DNN listed in Table~\ref{t:dnns}. We can see that DS2 and GNMT have a similar distribution of unique weights centered around $34$, while the rest have a more dispersed distribution centered around $50$. In all the cases, all the input neurons have less than 128 unique weights. Table~\ref{t:avg_uw_ruw} shows the average number of unique weights per input neuron (\textbf{UW/I}). On average, each input neuron from our set of DNNs has $44$ unique weights when applying an 8 bit quantization, out of the $256$ potential weights per input. Furthermore, Table~\ref{t:avg_uw_ruw} shows the percentage of multiplications of inputs by unique weights (\textbf{MULs}). As it can be seen, only between 0.57\% and 3.77\% of the multiplications are required to compute the FC layers.

\begin{table}[t!]
\begin{center}
\begin{tabular}{ccc}
\hline
\textbf{DNN Model} & \textbf{UW/I} & \textbf{MULs (\%)} \\
\hline
    DS2 & 38 & 1.67 \\
    GNMT & 29 & 0.57 \\
    Transformer & 49 & 3.77 \\
    Kaldi & 59 & 2.95 \\
    PTBLM & 43 & 0.71 \\
\hline
\end{tabular}
\end{center}
\caption{\textbf{UW/I} shows the average number of unique weights per input neuron. \textbf{MULs} is the percentage of multiplications of inputs by unique weights with respect to the total number of multiplications in the original model.}
\vskip -0.20in
\label{t:avg_uw_ruw}
\end{table}

\section{Partial Product Reuse}\label{s:partial_reuse}
This section describes how the low number of unique weights per input in FC layers, characterized in Section~\ref{s:WR}, can be exploited for efficient DNN inference. We first present the partial product memoization mechanism (Section~\ref{s:partial_product_mem}), which saves multiplications and storage by leveraging repeated weights in each input, for any FC layer of the network. We then present an optimization, called partial product approximation (Section~\ref{s:partial_product_approx}), to further reduce multiplications and storage at the expense of a minor accuracy loss.

\subsection{Partial Product Memoization}\label{s:partial_product_mem}
Our computation reuse method consists of two main steps. First, it detects statically the unique weights of each input for all the FC layers after quantization. Then, it dynamically performs the partial products between the inputs and their unique weights, and memoizes the partial results. These partial products will be retrieved later to perform the final accumulations of each output neuron. We refer to this scheme as Computation Reuse and Efficient Weight Storage (CREW).

\begin{figure*}[ht!]
    \centering
    \subfloat[Partial Product Memoization]{
        \includegraphics[width=0.70\columnwidth]{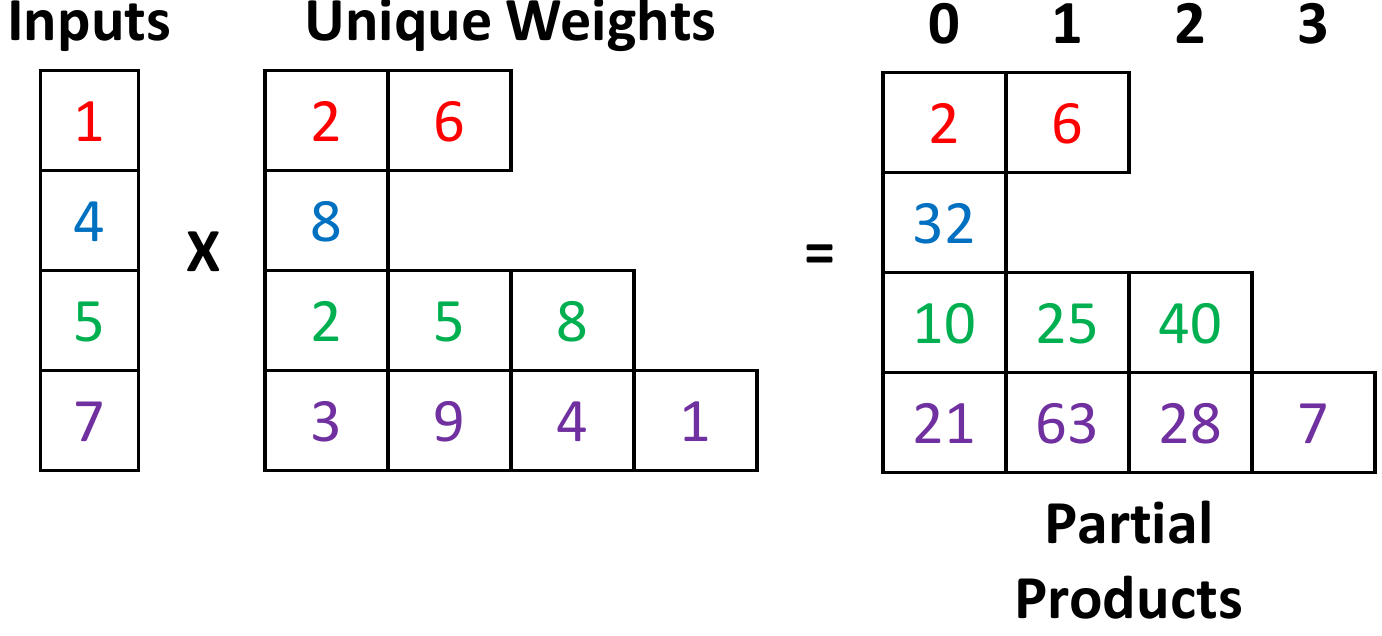}
        \hspace{1.0cm}
        \includegraphics[width=0.70\columnwidth]{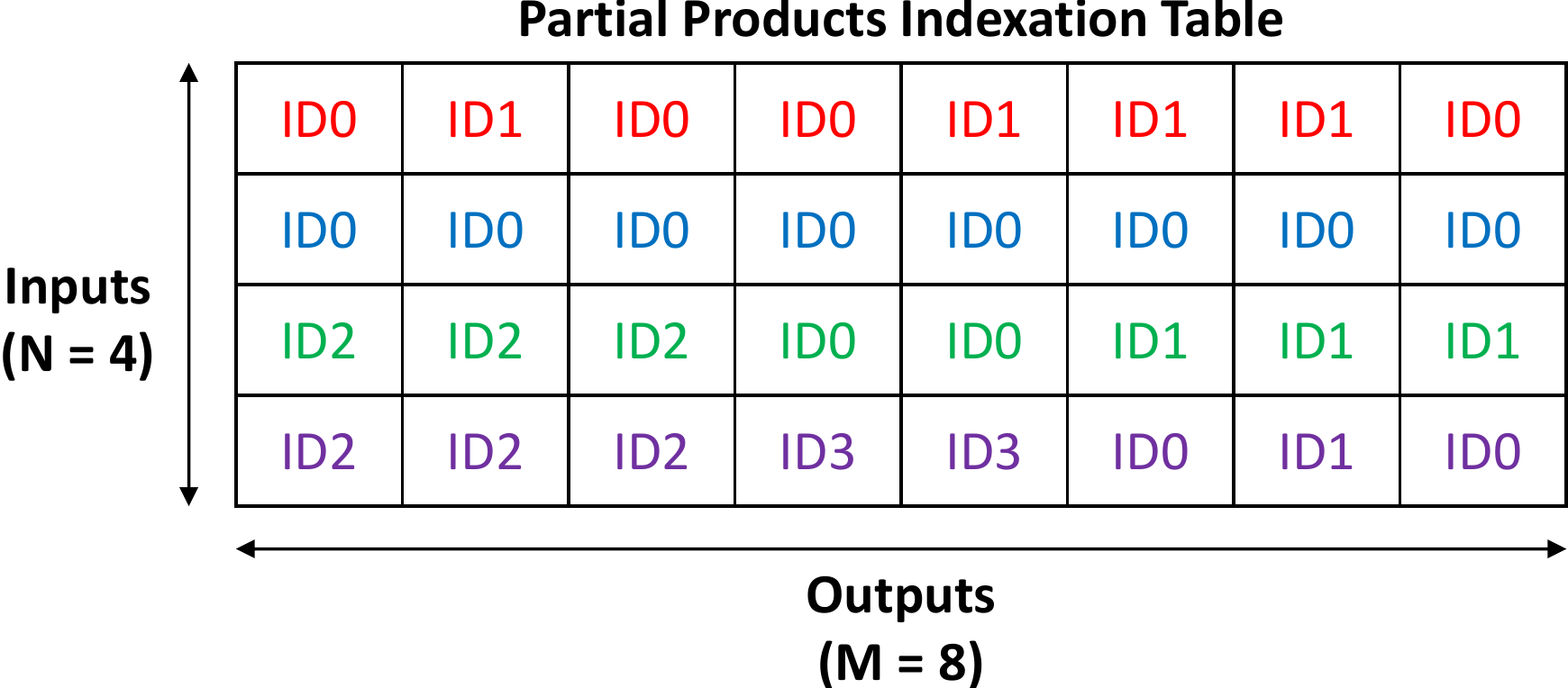}
        \label{f:pp_mem}
    } \\
    \subfloat[Partial Product Approximation]{
        \includegraphics[width=0.70\columnwidth]{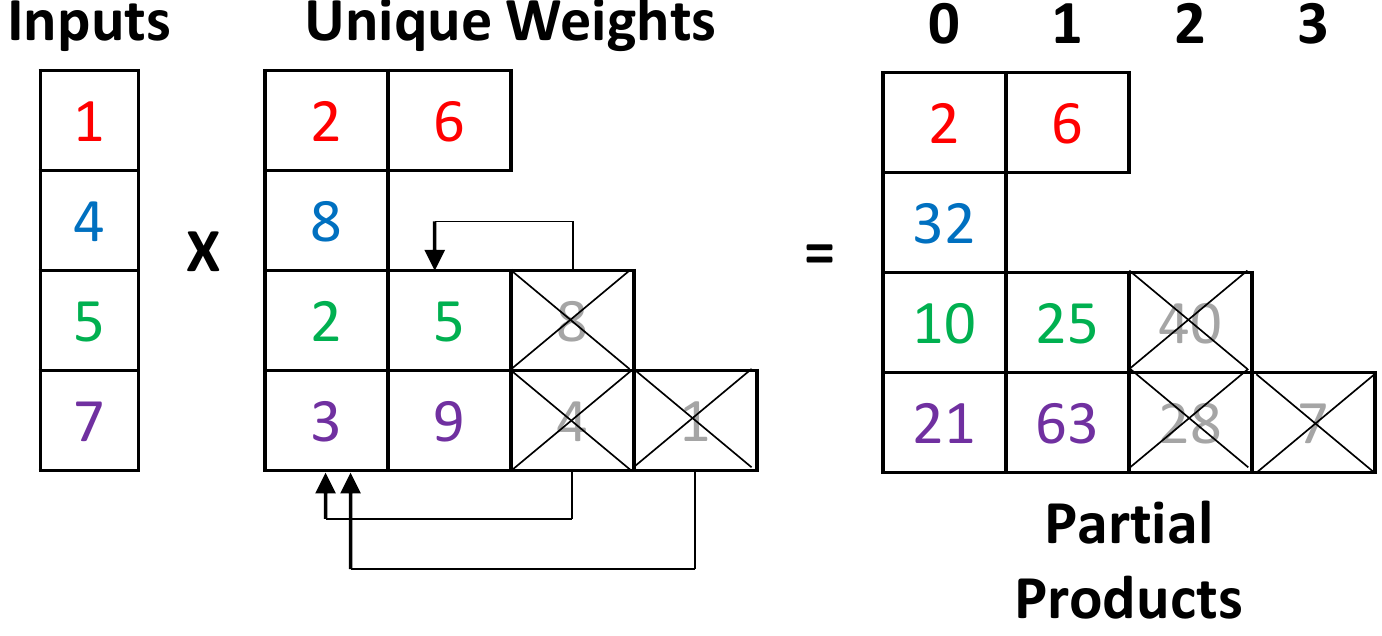}
        \hspace{1.0cm}
        \includegraphics[width=0.70\columnwidth]{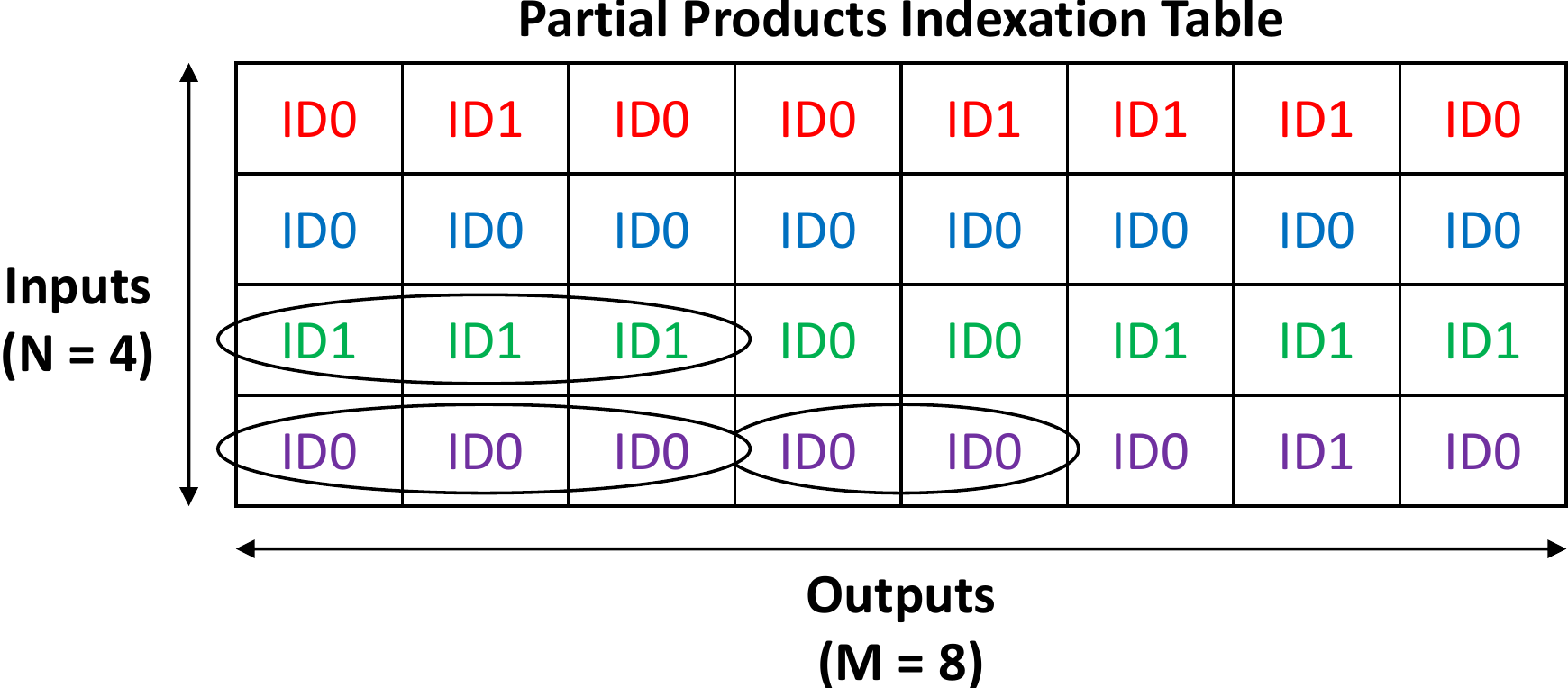}
        \label{f:approx_pp_mem}
    }
    \caption{Example of partial product memoization (a) vs partial product approximation (b) using a small FC layer.}
    \vskip -0.15in
    \label{f:crew_example}
\end{figure*}

The first step of the proposed approach is to detect the unique weights per input on each FC layer. This step can be done offline since the weights are statically known after training. The unique weights will be stored in a table in the same order as the inputs, that is, the unique weights of the first input will be stored first, next the unique weights of the second input and so on. There is an extra table storing the number of unique weights of each input. Then, we generate a table with the corresponding indexes to the partial products. Note that we are exploiting the partial products of repeated weights in the same spatial position along the $M$-axis as shown in Figure~\ref{f:fc_comp}. Therefore, the positions of the partial products are statically determined. Each FC layer will have an index table with $N * M$ entries. The indexes of each row $N$ may have a different bit length depending on the number of unique weights of the corresponding input. For example, if the number of unique weights of a given input neuron is 45, then the corresponding row of the table will have $M$ indexes of 6 bits.

The second step of CREW is applied during the dynamic execution of the DNN inference. Each FC will be computed as follows: first the multiplications of each input by their unique weights will be performed and the results memoized into a table of partial products. Then, indexes generated offline will be used to access the corresponding partial products of each output neuron, which will be accumulated to generate the final output. Note that each column of the table of indices contains $N$ indexes to a partial product related with a different input so that by adding them all, the output of each neuron is computed. Figure~\ref{f:pp_mem} illustrates the tables required by CREW for a small FC of $4 \times 8$ neurons. Note that in the example, the number of unique weights per input is $\leq 4$, as well as the number of partial products, so the size of the indices is $\leq 2$ bits.

\begin{table}[t!]
\begin{center}
\begin{small}
\begin{tabular}{ccc}
\hline
\textbf{DNN Model} & \textbf{Saved MULs (\%)} & \textbf{Storage Reduction (\%)} \\
\hline
    DS2 & 98 & 27 \\
    GNMT & 99 & 34 \\
    Transformer & 96 & 22 \\
    Kaldi & 97 & 16 \\
    PTBLM & 99 & 26 \\
\hline
\end{tabular}
\end{small}
\end{center}
\vskip -0.05in
\caption{Reduction in multiplications and storage.}
\vskip -0.15in
\label{t:avg_mul_comp}
\end{table}

One of the main benefits of the memoization of partial products is the reduction in multiplications. The number of multiplications per input is reduced to the number of unique weights it has. Moreover, the storage and memory accesses are also reduced since the indexes are smaller than the original weights. Table~\ref{t:avg_mul_comp} shows the multiplications and storage reduction achieved on a set of state-of-the-art DNNs. On average, we can reduce the storage required for the FC layers of the models by 25\% over the quantized networks (taking into account the required metadata for CREW), and save 98\% of the multiplications. CREW requires extra metadata such as the number of unique weights per input and the table of partial products. However, since the unique weights represent a small fraction of the total number of weights in the original model ($< 4\%$) and the indices are also smaller than the original weights, the extra storage required for metadata is negligible compared to the savings provided by CREW.

\subsection{Partial Product Approximation}\label{s:partial_product_approx}
\begin{figure}[t!]
\centering
\includegraphics[width=1\columnwidth]{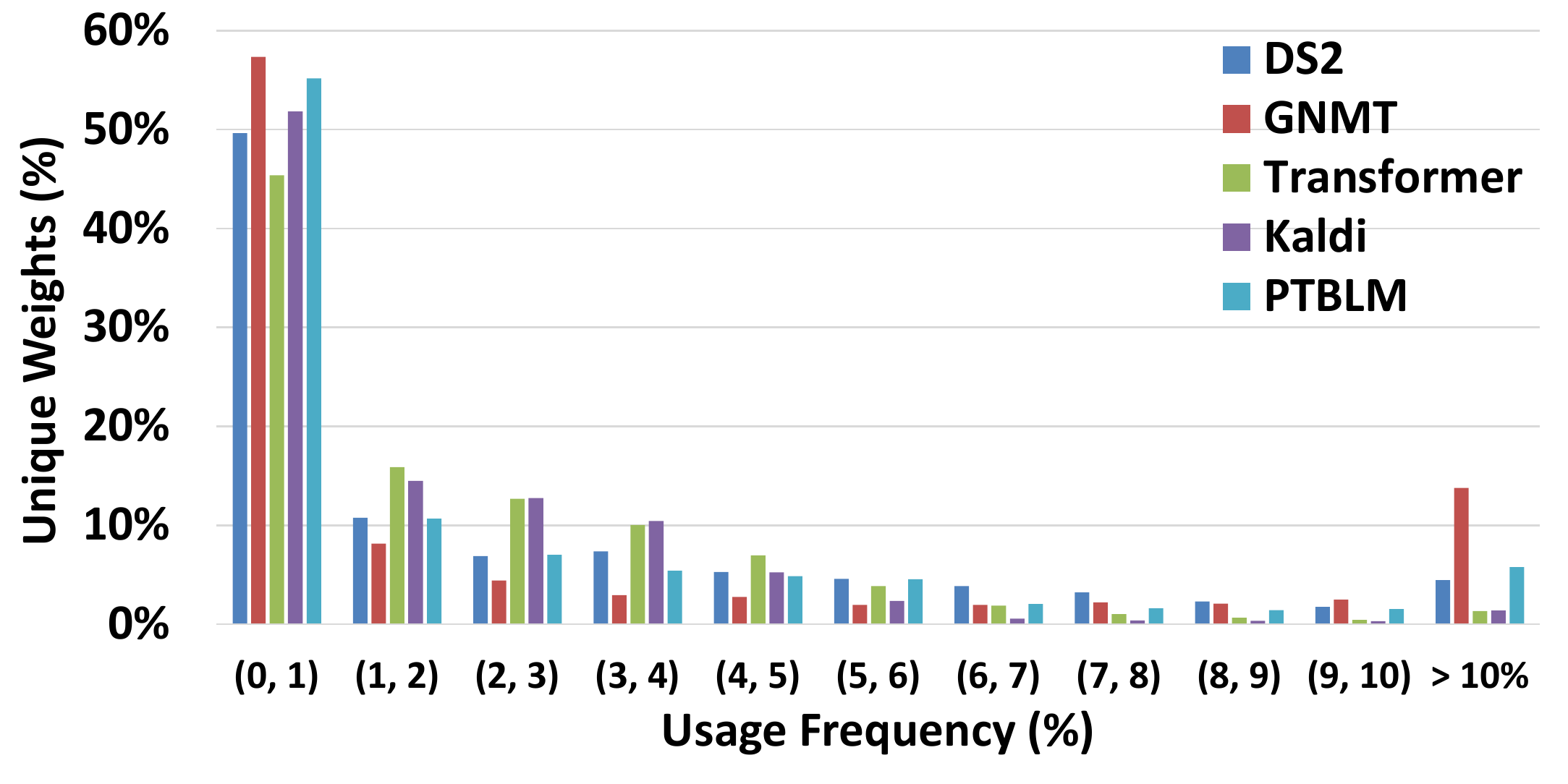}
\vskip -0.05in
\caption{Usage frequency histograms of the unique weights per input neuron for all the FC layers of DS2, GNMT, Transformer, Kaldi and PTBLM. The frequency of use is the number of times each unique weight is repeated divided by the total number of weights for each input neuron.}
\vskip -0.15in
\label{f:all_uw_freq_hist}
\end{figure}

Partial product memoization reduces multiplications by exploiting weight repetition. The original weights are replaced by the unique weights and the indexes to the partial products. The number of unique weights per input determines the size of the indexes and, hence, the total size of the DNN. We propose to further improve our mechanism by reducing the number of unique weights. From the histograms of the unique weights frequency usage shown in Figure~\ref{f:all_uw_freq_hist}, we observe that on average more than 50\% of the unique weights have a frequency of use lower than 1\%. Based on this observation we propose to approximate the less frequent unique weights by its closest value in the remaining unique weights.

The partial product approximation technique reduces the number of unique weights of some (potentially all) input neurons to the nearest and smaller power of two. For this purpose, a subset of unique weights, as well as their respective partial products, are approximated by similar unique weights/products of the same input. By reducing the number of unique weights to a lower power of two, the index bit length can be reduced. For example, if an input neuron has 38 unique weights, we can approximate 6 of them so that the number of unique weights becomes 32, reducing all the indexes related to this input by one bit. Figure~\ref{f:approx_pp_mem} shows the approximation of three unique weights from two different input neurons that result in a reduction of one bit in the corresponding indexes of the index table. As a result of this optimization, all the indexes of the example require only one bit.

\begin{algorithm}[t!]
\scriptsize 
\caption{Partial Product Approximation Heuristic}
\begin{algorithmic} [t]
    \STATE $W = LoadWeights()$;
    \STATE $Thr = Constant$;
    \FOR {Input Neuron ($i$)}
        \STATE $UW, Frequency = UniqueWeights(W[i], counts=True)$;
        \STATE $Current2Power = 2^{(log_2 UW)}$;
        \STATE $Low2Power = Current2Power/2$;
        \STATE $dist\_W = UW - Low2Power$;
        \STATE $F = Sort(Frequency)$;
        \STATE $num\_weights = Size(W[i])$;
        \STATE $low\_freq\_W, del\_uw = LowFreqSum(UW, F, dist\_W)$;
        \STATE $WR = (low\_freq\_W/num\_weights)$;
        \IF {$WR < Thr$}
            \STATE $sim\_uw = ClosestWeights(UW, del\_uw)$;
            \STATE $new\_W = ReplaceApproximate(W, sim\_uw, del\_uw)$;
        \ENDIF
    \ENDFOR
    \STATE $Accuracy = Inference(new\_W)$;
\end{algorithmic}
\label{alg:approx_pp_mem}
\end{algorithm}

The reduction in storage and memory accesses obtained from this optimization may affect accuracy if it is not carefully applied. There is a trade-off between the number of weights that can be approximated without accuracy loss (or with a negligible loss) and the benefits obtained in terms of performance and energy consumption during DNN inference. The heuristic used to select which weights are approximated is shown in Algorithm~\ref{alg:approx_pp_mem}. First, for each input neuron we compute the number of weights to be approximated by measuring the distance ($dist\_W$) between the number of unique weights ($UW$) and the closest smaller power of two ($Low2Power$). Then, we select the weights to approximate by sorting the frequencies from lowest to highest, that is the number of times each unique weight is repeated in the original model. The rationale is that approximating the weights that are less frequently used should introduce the smallest distortion in the model. The frequencies of the $dist\_W$ lowest weights are added ($low\_freq\_W$) to determine the relevance of the weights that will be approximated ($WR$), and the result is compared against a threshold ($Thr$): if the total frequencies of the less common weights ($WR$) is smaller than the threshold ($Thr$) then the weights are considered to have a small importance in the layer and are selected for approximation; otherwise, approximation is not used for this input. Finally, the selected weights ($del\_uw$) are replaced in the original model by their closest unique weight ($sim\_uw$) and the new model accuracy is tested. This heuristic can be generalized to reduce more than one bit by shrinking the number of unique weights to different powers of two while fulfilling the threshold condition.

\begin{figure}[t!]
\centering
\includegraphics[width=1\columnwidth]{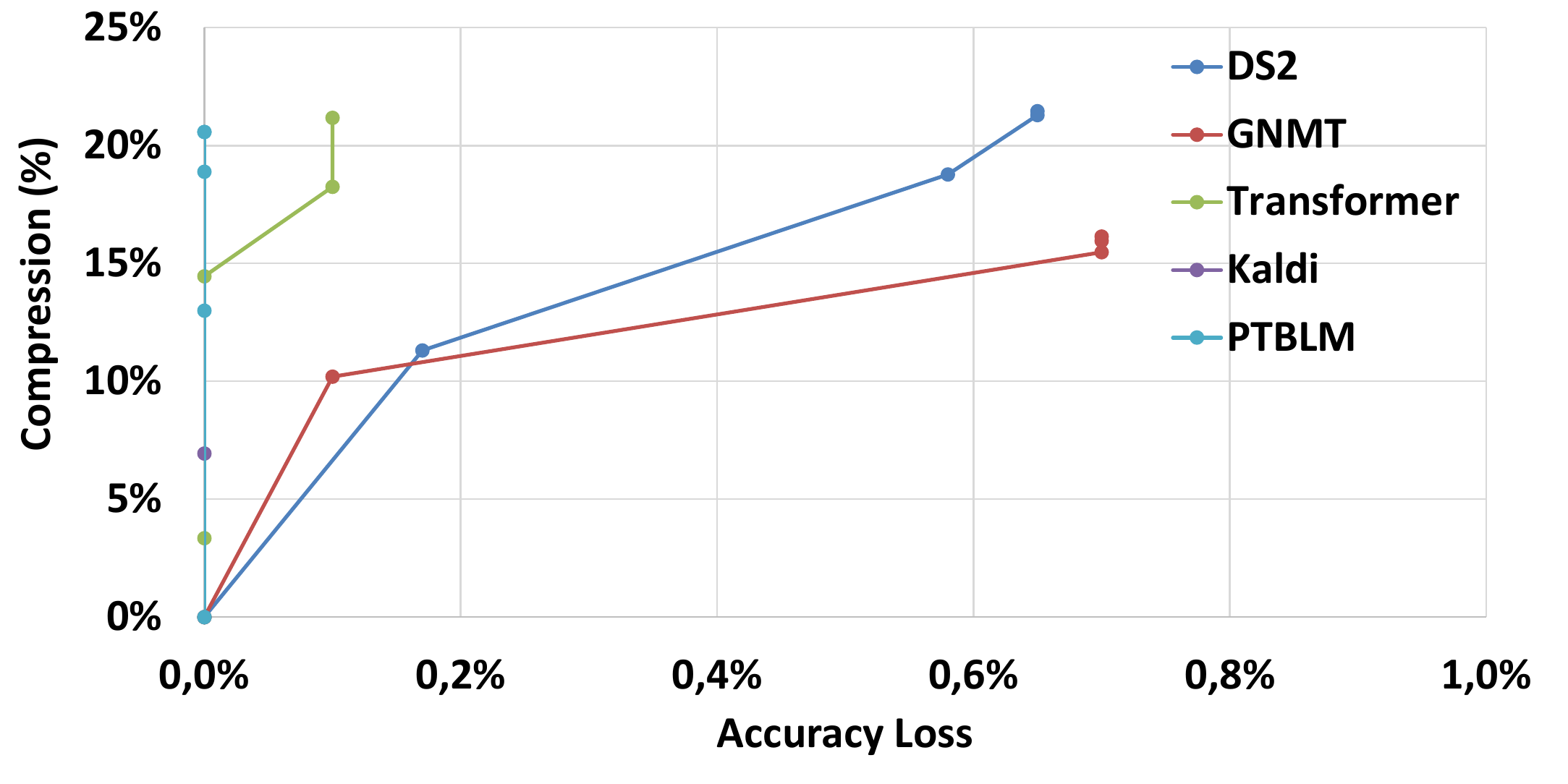}
\caption{Accuracy loss versus compression ratio over partial product memoization (CREW without approximation) for different thresholds of the heuristic for unique weight approximation. Thresholds tested go up to 20\% in steps of 5\%, where the initial 0\% threshold means no weight approximation.}
\vskip -0.20in
\label{f:sens_thr}
\end{figure}

A small threshold will limit the benefits achieved since less input neurons will be able to reduce their number of unique weights. On the other hand, a high threshold value may negatively impact the DNN accuracy since important weights may be approximated. We performed a sensitivity analysis on a representative set of DNNs to determine a proper threshold. Figure~\ref{f:sens_thr} shows that we can achieve, on average, an extra 17\% model compression over our partial product memoization mechanism (CREW without accuracy loss as described in Section~\ref{s:partial_product_mem}) while losing less than 1\% of accuracy in absolute terms. We observed that with a threshold of 10\%, the approximation covers more than 90\% of the input neurons, which means that 90\% of the indices are being reduced by one bit. For the Transformer and PTBLM networks the accuracy loss is almost negligible, so we tested a more aggressive approximation trying to reduce up to 2 bits per index. By doing so, we achieved an extra 35\% model compression while losing less than 1\% of accuracy. In summary, we can achieve extra improvements in performance and energy consumption during DNN inference in scenarios where the user is willing to accept a very minor accuracy loss.

\section{CREW Accelerator}\label{s:accelerator}
This section describes the hardware support required to implement CREW. First, we present the main hardware components of the CREW accelerator for DNN inference. Next, we describe how FC layers are executed in the accelerator using CREW with an enhanced weight stationary dataflow. Finally, we describe how to efficiently support other types of layers and networks such as CNNs.

\begin{figure}[t!]
\centering
\includegraphics[width=0.85\columnwidth]{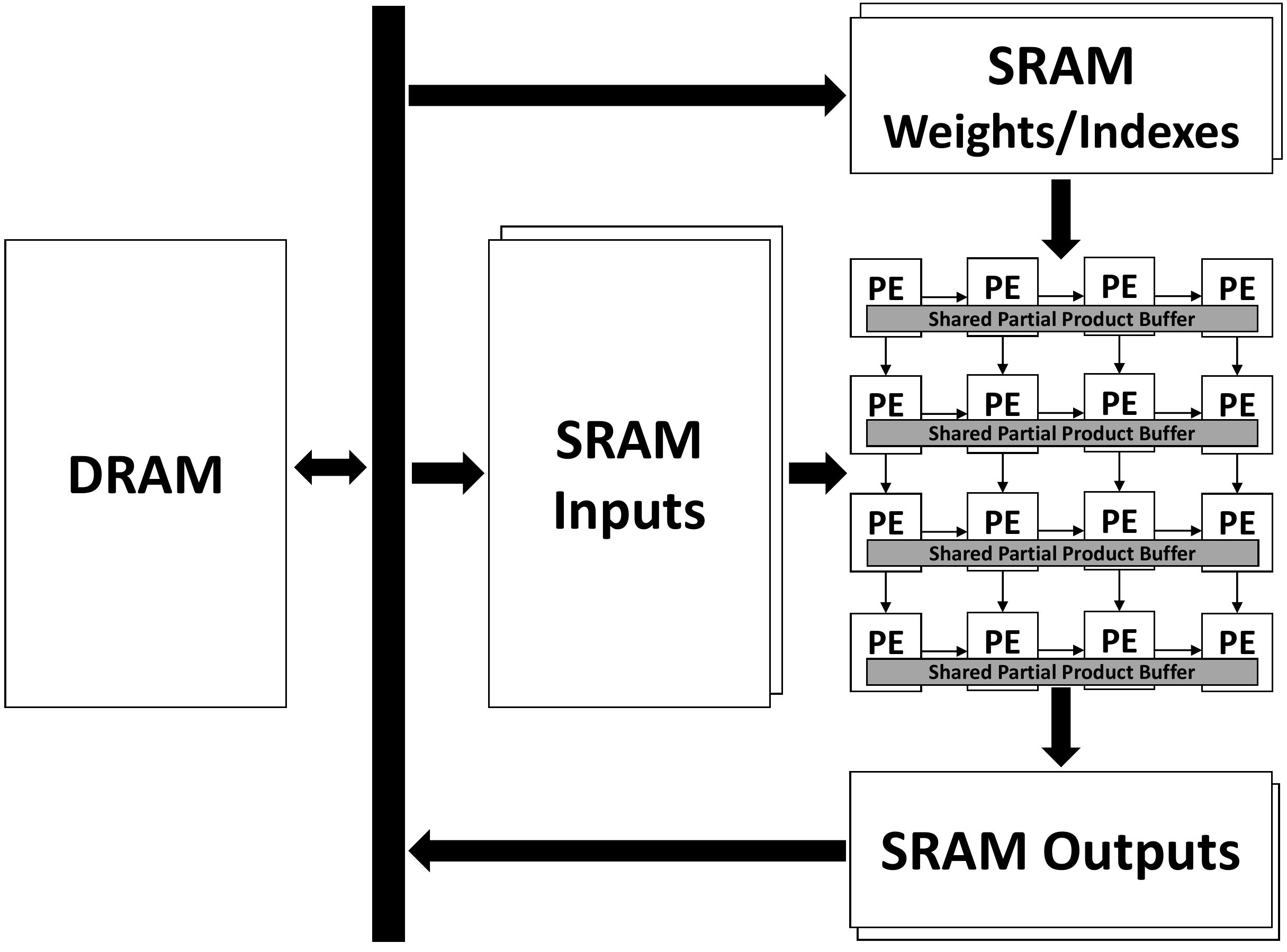}
\caption{Architecture of the CREW accelerator.}
\vskip -0.15in
\label{f:crew_accelerator}
\end{figure}

\begin{figure}[t!]
\centering
\includegraphics[width=1.0\columnwidth]{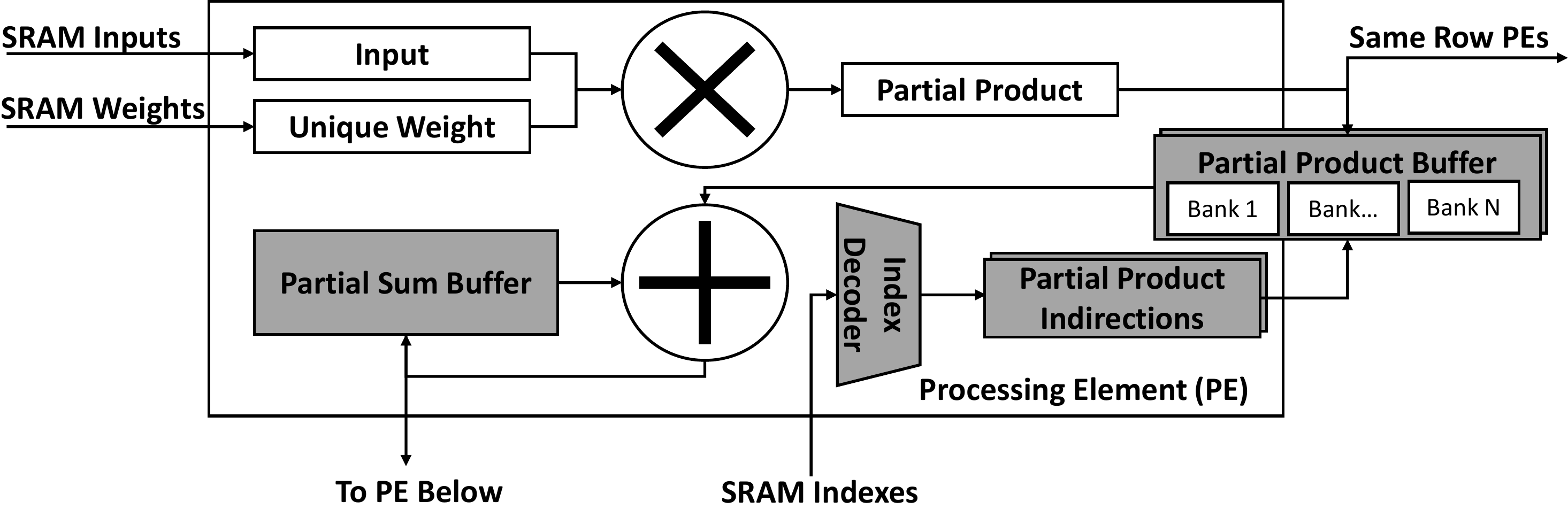}
\caption{Processing Element (PE) Architecture. Components shaded in gray represent the required extra hardware for CREW. The partial product buffer is shared among all or a subset of PEs of the same row of the array.}
\vskip -0.15in
\label{f:crew_pe}
\end{figure}

\begin{figure*}[t!]
\centering
\includegraphics[width=1.50\columnwidth]{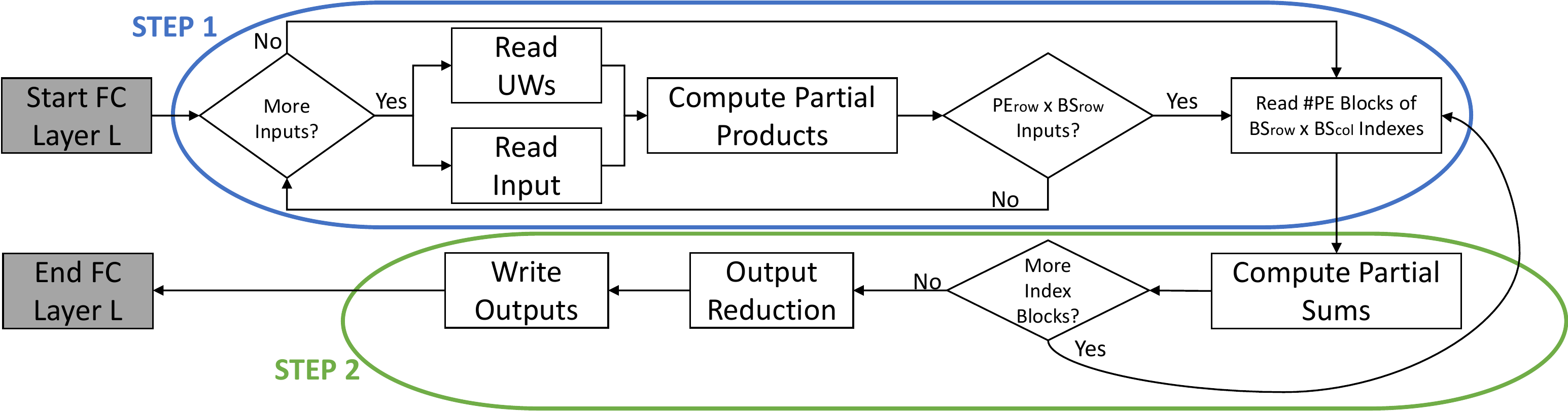}
\caption{CREW Execution Dataflow. $PE_{row}$ is the number of PEs per row in the systolic array, and $BS_{row}$ and $BS_{col}$ determine the block size, i.e. the number of indexes per block. $BS_{row}$ refers to the number of indexes relative to different input neurons, while $BS_{col}$ refers to indexes associated to the same input neuron but used in the computation of different output neurons.}
\vskip -0.15in
\label{f:dataflow}
\end{figure*}

\subsection{Architecture}\label{s:crew_architecture}
In this section, we present an accelerator that takes advantage of the CREW mechanism to reuse computations in the FC layers of different DNNs. CREW exploits the high degree of repeated weights at each input neuron to save computations and memory accesses. Figure~\ref{f:crew_accelerator} shows a high-level schematic of the architecture. The main components are the blocks of SRAM used for the inputs, outputs, unique weights and indexes, and the systolic array of processing elements (PEs). Each PE includes functional units to perform computations. All the global SRAM memories are double buffered to load data from main memory while performing computations, and highly multi-banked to achieve the bandwidth required to feed a large number of PEs. This systolic array architecture is inspired in Google's TPU~\cite{TPU}, which has proven to be an efficient accelerator to perform DNN inference.

Figure~\ref{f:crew_pe} shows the architecture of a Processing Element (PE). Each PE has a multiplier and an adder which can be used independently, that is, the multipliers are used to perform the unique weight multiplications and can be power-gated during the rest of the execution. On the other hand, the adders are used during the entire execution to accumulate the partial products of each output into the Partial Sum Buffer. Each PE requires additional memory buffers to store the partial products and the corresponding indexes to access them. The indexes are decoded from the compressed stream and padded to 8 bits to be stored in the partial product indirections buffer. The partial products are stored in a multi-banked buffer shared among all the PEs from the same row of the array.

\subsection{Dataflow}\label{s:crew_dataflow}
We implement an enhanced weight stationary dataflow to provide better efficiency in the execution of the FC layers with CREW. Traditional DNN accelerators follow one of these dataflows: output stationary, weight stationary or input stationary. In output stationary, each PE computes an output at a time. In the weight/input stationary dataflow, each PE pre-loads a weight/input from memory to its local register, and those are used to perform all the associated computations. FC layers do not tend to be efficiently executed on systolic arrays because the potential reuse is much more limited compared to convolutional layers. Given a batch size of one, only the inputs can be reused multiple times to compute each output, making all the common dataflows inefficient to exploit the resources of the systolic array. We propose to use an enhanced weight stationary dataflow coupled with a blocking scheme.

The CREW accelerator executes FC layers in two main steps carried out in parallel as is shown in the flowchart of Figure~\ref{f:dataflow}, where both steps are marked in different colors. In the first step, marked in blue, the accelerator starts by reading an input and its number of unique weights. Then, the unique weights of the input are read from memory. The input is broadcasted to a row of PEs and the unique weights are distributed along the PEs of the same row. Each PE will multiply the input and its unique weights and store the partial results into the Partial Product Buffer shared by all the PEs of a given row. This step is repeated until all the partial products of unique weights for the FC layer are computed. Note that the multipliers can be power gated after doing all the unique multiplications.

As it is shown in the schematic of Figure~\ref{f:dataflow}, when a given number of partial products is stored in the shared buffer of each row of PEs, the second step starts in parallel with the first one. The number of partial products required is determined by a block size parameter of $BS_{row} \times BS_{col}$ so that when the unique weights of $BS_{row}$ inputs have been processed for each row of the systolic array ($PE_{row}$), the second step begins. The block size also determines the number of indexes ($BS_{row} \times BS_{col}$) stored in the index table of each PE and the number of partial outputs ($BS_{col}$) that each PE is computing. During the second step, the accelerator reads one block of $BS_{row} \times BS_{col}$ indexes for each PE to perform the partial sums, as illustrated by the green marker in Figure~\ref{f:dataflow}. The blocks of indexes are constructed offline and stored consecutively in main memory. Each PE will receive a block of indexes pointing to partial products stored in its shared buffer. Therefore, each PE in a given row of the array is computing a different set of partial outputs, while the PEs in the same column are computing different partial sums for the same set of outputs.

\begin{figure*}[t!]
    \centering
    \subfloat[Blocking Scheme]{
        \includegraphics[width=1.00\columnwidth]{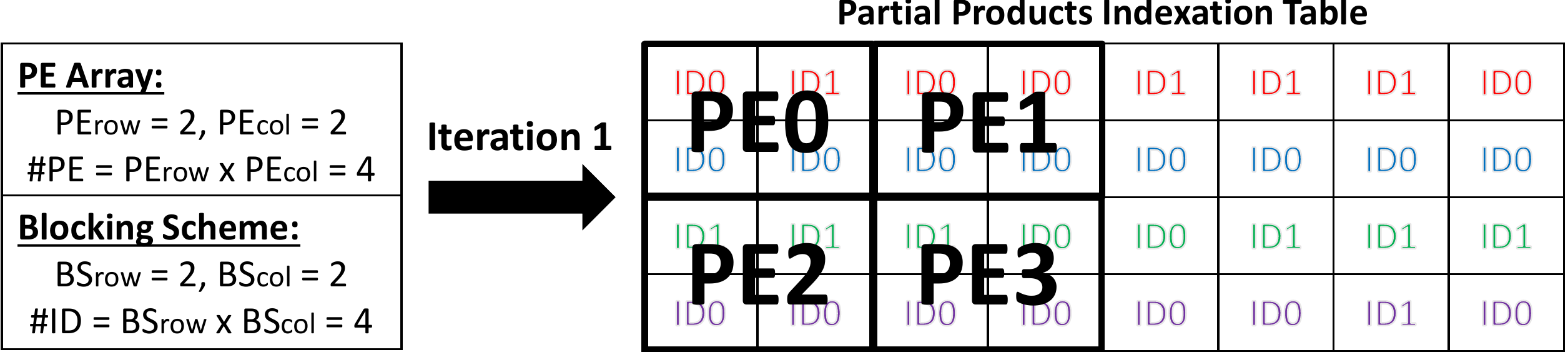}
        \includegraphics[width=0.75\columnwidth]{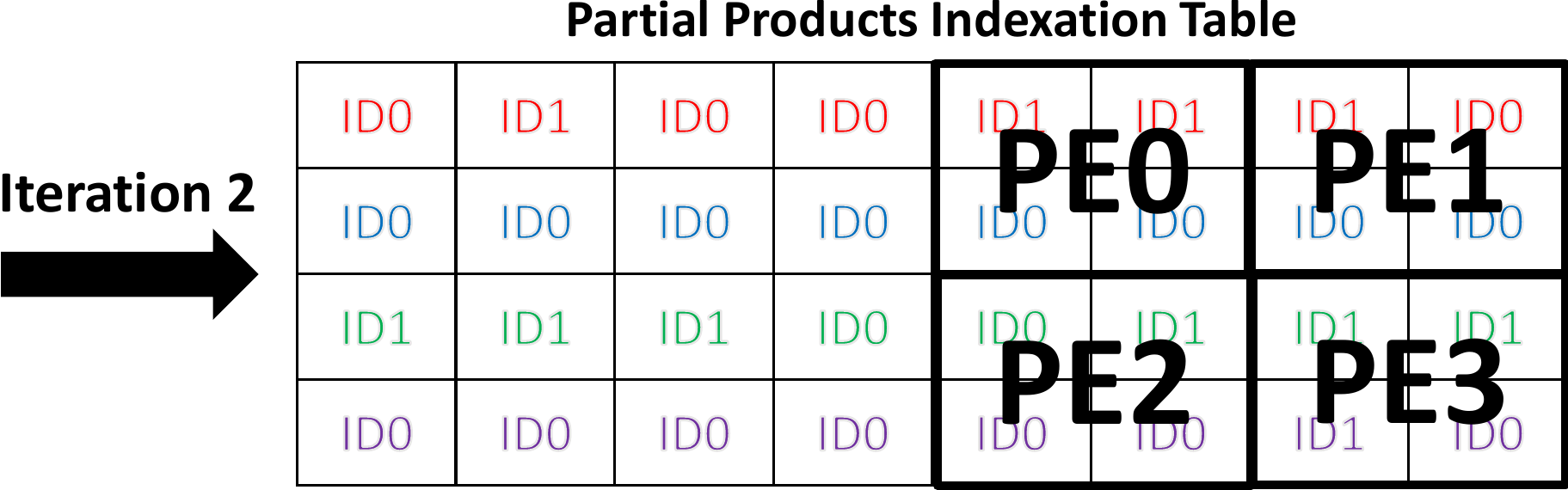}
        \label{f:blocking_example}
    } \\
    \subfloat[FC Execution Example]{
        \includegraphics[width=1.75\columnwidth]{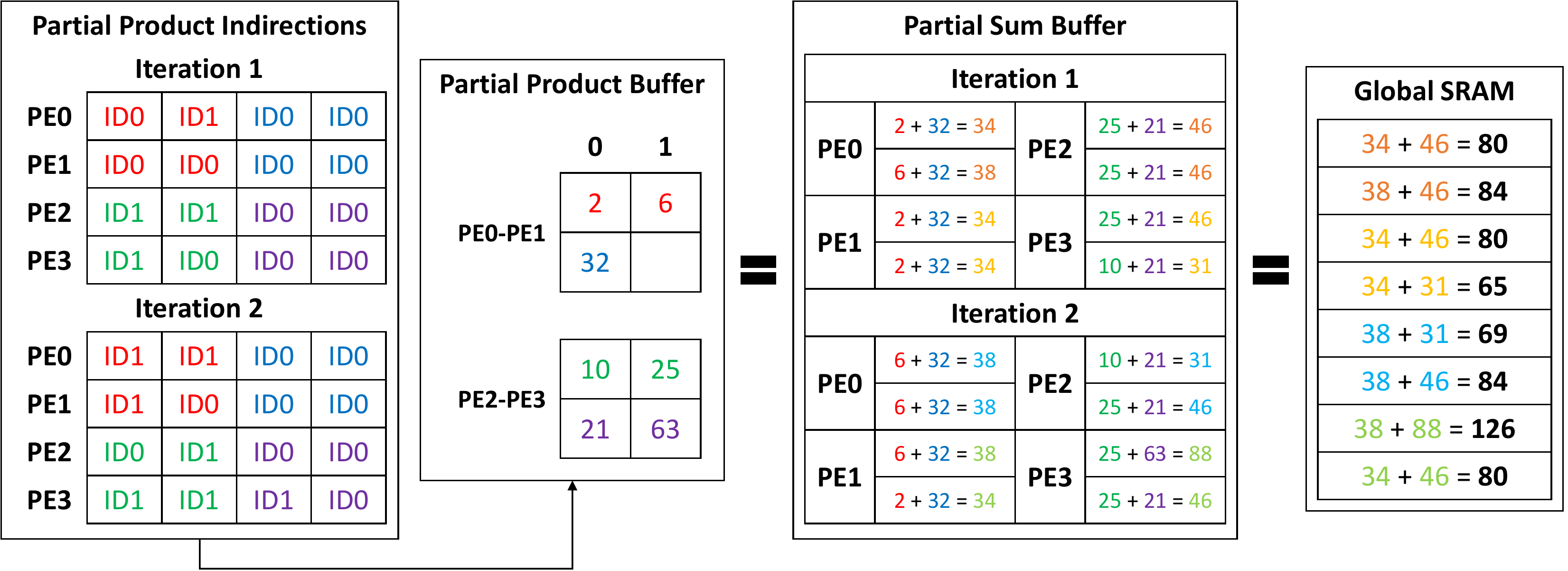}
        \label{f:dataflow_example}
    }
    \caption{FC Execution in the CREW Accelerator.}
    \vskip -0.15in
    \label{f:fc_execution}
\end{figure*}

The indexes of a block arrive in a compressed format to each PE, that is, all the $BS_{col}$ indexes of a given row of the block have the same size, but for each $BS_{row}$ the size of the indexes may be different. The reasoning behind the compressed format is that the blocks are composed of indexes related to different input neurons which may have a different number of unique weights, and so indexes exhibit variable size. All the indexes of a block are stored consecutively in memory starting by the $BS_{col}$ indexes of the first row of the block. Each PE includes a specialized hardware decoder to decompress the indexes and store them uncompressed (8 bits per index) in the Partial Product Indirections buffer. The number of indexes in a block is fixed and known by the control unit of each PE. Similarly, the sizes of the indexes related to each input neuron are computed offline, stored in memory, and sent along the indexes of each block when needed. Note that a single value of three bits per input neuron is enough to determine the sizes of all the indexes. The hardware required to perform the decoding of each index is relatively simple since all the information is statically available. The decoder will receive the compressed indexes of a block in chunks, the information of the size of the indexes of the block and the number of indexes in the block. Then, it will read a byte of the chunk, and use the information of the size of the first index to discard the unneeded bits and pad the index to 8 bits. Next, a pointer will be increased using the size of the index to point at the start of the next index of the chunk. The decoder also takes into account how many indexes of a given size are in the block, to use the corresponding information at each decoding. The decoding of the indexes is overlapped with the loading and distribution of the blocks as well as the computations.

On the other hand, the partial products between inputs and unique weights can be computed in parallel to the partial sums, along with the loading and decoding of the next blocks of indexes from the global buffer. To this end, the shared Partial Product Buffer and the indirection table of each PE are double buffered. To avoid collisions, the shared buffer of partial products is highly multi-banked and each PE starts to access partial products relative to input neurons with a different offset located into a different bank. More specifically, the shared buffer of partial products for a given row of the array is modeled to have a bank per $PE_{col}$. Then, each bank is sized to store the partial products related to $BS_{row}/PE_{col}$ input neurons. The worst case scenario is considered, so that up to $256$ partial products may be stored for each input neuron when using an 8-bit quantization. Note that in this case the partial products will have a bit-length of 16 bits.

Continuing the example from Section~\ref{s:partial_reuse}, Figure~\ref{f:fc_execution} shows an FC execution in the CREW accelerator following our dataflow and blocking scheme. Considering a systolic array of $2 \times 2$ PEs and a block size of $4$ indexes, the blocks of indexes are distributed in two iterations as shown in Figure~\ref{f:blocking_example}. Figure~\ref{f:dataflow_example} illustrates the second step of our dataflow, having all the partial products computed in the first step stored in the shared buffer of each row of PEs, and all the decoded indexes in the indirections buffer. At each cycle, each PE reads an index of its block, the associated partial product and the current partial sum of the corresponding output to perform the next accumulation. The partial sum buffer depicted in the figure shows the additions performed by each PE in each iteration. Once the PE has processed all the indexes of the block, it proceeds to compute the subsequent block. This step will be repeated until all the blocks are processed. Finally, a reduction of all the partial summations is performed from top to bottom of the systolic array, writing the final results into the SRAM output buffer as shown in the example.

In summary, each PE will compute partial sums of multiple neurons by accessing to different partial products associated to the same set of inputs. The two main steps are repeated until all the partial products of each input are computed and all the indexes processed. Note that the partial products are computed only once and reused during the rest of the FC layer execution.

\subsection{Design Flexibility}\label{s:crew_flexibility}
A key parameter in the dataflow of CREW and its hardware implementation is the block size. The block size affects the performance and the size of the local buffers in each PE. A large block size will require sizable local buffers in each PE. On the other hand, the block size has to be large enough to be able to perform computations while loading from memory the next blocks of indexes, in order to hide main memory latency, and avoid collisions to the shared buffers. To avoid large interconnection overheads in case of a considerable number of columns in the systolic array, the shared buffers can be replicated and shared among PEs of a given subset for the same row of the array. Therefore, by adjusting the block size and resizing the buffers, CREW can keep the scalability of the accelerator to any systolic array size. Although our accelerator is specialized in efficient inference of FC layers by using CREW, it can support any kind of layer such as convolutionals as executed in a TPU-like accelerator, without obtaining the benefits of the reuse mechanism.

\section{Evaluation Methodology}\label{s:methodology}
We have extended ScaleSim~\cite{scalesim}, a simulator for DNN accelerators, to accurately model three different systems: a TPU-like architecture, UCNN~\cite{UCNN} and our CREW scheme presented in Section~\ref{s:accelerator}. ScaleSim models a systolic array architecture and supports any kind of neural network, including CNNs, MLPs and RNNs. Table~\ref{t:hw_params} shows the parameters for the experiments. The baseline configuration is a TPU-like architecture clocked at 500 MHz with an output stationary dataflow~\cite{Eyeriss}. It includes a systolic array of $16 \times 16$ (256 PEs) and 24 MB of global on-chip SRAM. Inputs and weights are quantized to 8-bit integers without any accuracy loss for our set of DNNs, whereas activation functions are performed in 32-bit floating point. On top of this architecture, we have implemented UCNN as described in~\cite{UCNN} and our CREW scheme as presented in Section~\ref{s:accelerator}. The sizes of the additional local buffers required by UCNN and CREW are provided in Table~\ref{t:hw_params}. CREW requires three local SRAM buffers on each PE: the Partial Product Indirections buffer, the Partial Product Buffer shared between multiple PEs of the same row and the Partial Sum Buffer. All the local buffers are sized taking into account the block size parameter of our dataflow. We employ a block size of $16 \times 16$ for all the DNNs, as we found that it provides a good trade-off between on-chip storage requirements and performance. As a result, all the local buffers require less than 1KB of capacity per PE. These overheads are taken into account for the area, timing and energy evaluation of the accelerator. Regarding main memory, we model an LPDDR4 of 8 GB with a bandwidth of 16 GB/s (dual channel).

\begin{table}[t!]
\centering
\resizebox{0.75\columnwidth}{!}{
    \centering
    \begin{tabular}{|c|c|}
    \hline
    \multicolumn{2}{|c|}{\cellcolor[gray]{0.9} \small \textbf{Common Parameters}}\\
    \hline
    \small Technology&\small 32 nm\\
    \small Frequency&\small 500 MHz\\
    \small \# PEs& \small 16 x 16\\
    \small Block Size& \small 16 x 16\\
    \small Total Global SRAM Buffers Size& \small 24 MB\\
    \hline
    \multicolumn{2}{|c|}{\cellcolor[gray]{0.9} \small \textbf{UCNN Parameters}}\\
    \hline
    \small Input/Weight Indirections Buffer Size/PE& \small 0.75 KB\\
    \small Input/Weight Buffer Size/PE& \small 272 B\\
    \small Partial Sum Buffer Size/PE& \small 0.75 KB\\
    \hline
    \multicolumn{2}{|c|}{\cellcolor[gray]{0.9} \small \textbf{CREW Parameters}}\\
    \hline
    \small Partial Product Indirections Buffer Size/PE& \small 0.5 KB\\
    \small Partial Product Buffer Size/PE& \small 1 KB\\
    \small Partial Sum Buffer Size/PE& \small 0.75 KB\\
    \hline
    \end{tabular}
}
\vskip -0.05in
\caption{Parameters for the accelerators.}
\vskip -0.10in
\label{t:hw_params}
\end{table}

Regarding area and energy consumption, the combinational logic hardware is implemented in Verilog, including all the additional components required by CREW, and synthesized to obtain the delay and power using the Synopsys Design Compiler, the modules of the DesignWare library and the technology library of 28/32nm from Synopsys~\cite{Synopsys}. For the technology library, we use the standard low power configuration with 0.78V. On the other hand, we characterize the memory components of the accelerator by obtaining the delay, energy per access and area using CACTI-P~\cite{Cacti}. We use the configurations optimized for low power and a supply voltage of 0.78V. Finally, the energy consumption of main memory is estimated by using the MICRON power model for LPDDR4~\cite{Micron}. The results obtained with the aforementioned tools are combined with the activity factors and memory traces provided by the extended ScaleSim~\cite{scalesim} simulator to obtain the dynamic and static power of the accelerators.

\begin{table}[t!]
\begin{center}
\begin{tabular}{ccc}
\hline
\textbf{DNN Model} & \textbf{Size (MB)} & \textbf{Accuracy} \\
\hline
    DS2 & 144 & 10.32\% (WER) \\
    GNMT & 518 & 29.8 (BLEU) \\
    Transformer & 336 & 28.0 (BLEU) \\
    Kaldi & 18 & 10.85\% (WER) \\
    PTBLM & 137 & 78.15 (Perplexity) \\
\hline
\end{tabular}
\end{center}
\vskip -0.10in
\caption{DNNs employed for the experimental evaluation of CREW. The model size accounts for the original FP parameters of the FC layers where CREW is applied.}
\vskip -0.20in
\label{t:dnns}
\end{table}

Our objective is to prove that our scheme provides important savings for multiple applications and different DNN architectures. To this end, we evaluate our technique on five state-of-the-art DNNs from different application domains, including speech recognition and machine translation as shown in Table~\ref{t:dnns}. We include DeepSpeech2 ($DS2$)~\cite{deepspeech2}, an RNN that consists of GRU cells for end-to-end speech recognition implemented in PyTorch~\cite{pytorch}, and Kaldi~\cite{Kaldi}, an MLP for acoustic scoring from the popular framework of the same name. DS2 and Kaldi are trained and evaluated with the Librispeech~\cite{panayotov2015librispeech} dataset. In addition, we employ two machine translation networks: $GNMT$~\cite{gnmt} and Transformer~\cite{attention}. GNMT is an LSTM network for neural machine translation based on the Google Translator and trained using the WMT16~\cite{WMT16} dataset with texts of newspapers from German to English (DE-EN). As described in Section~\ref{s:DNN}, the Transformer is a deep MLP with an encoder-decoder architecture that is mainly composed of attention layers. We evaluate the Transformer implementation from the OpenSeq2Seq~\cite{OpenSeq2Seq} framework of NVIDIA, using the WMT16 dataset for translating from English to German (EN-DE). Finally, PTBLM~\cite{PTBLM} is an LSTM network for language modeling using the Penn Treebank dataset. Accuracy is reported as Word Error Rate (WER) for speech recognition (lower is better), bilingual evaluation understudy (BLEU) for machine translation (higher is better), and perplexity for language modeling (lower is better). For all the DNNs, we employ the entire evaluation sets from their respective datasets, including several hours of audio and a large number of texts, to assess the efficiency of our computation reuse scheme in terms of performance and energy consumption. Our workloads represent important machine learning applications, and the selected DNNs cover three of the most common approaches for sequence processing: MLPs, LSTMs and GRUs.

\section{Results}\label{s:results}
This section evaluates the performance and energy efficiency of our computation reuse scheme. First, we present the speedups and energy savings achieved by CREW compared to a baseline TPU-like accelerator and UCNN. In order to do a fair comparison, we evaluate the factorization of UCNN using the same blocking dataflow as CREW. Next, we present the additional benefits of CREW when applying the partial product approximation presented in Section~\ref{s:partial_product_approx}. Finally, we discuss the accelerator overheads.

\subsection{CREW Evaluation}\label{s:crew_eval}
\begin{figure}[t!]
\centering
\includegraphics[width=0.75\columnwidth]{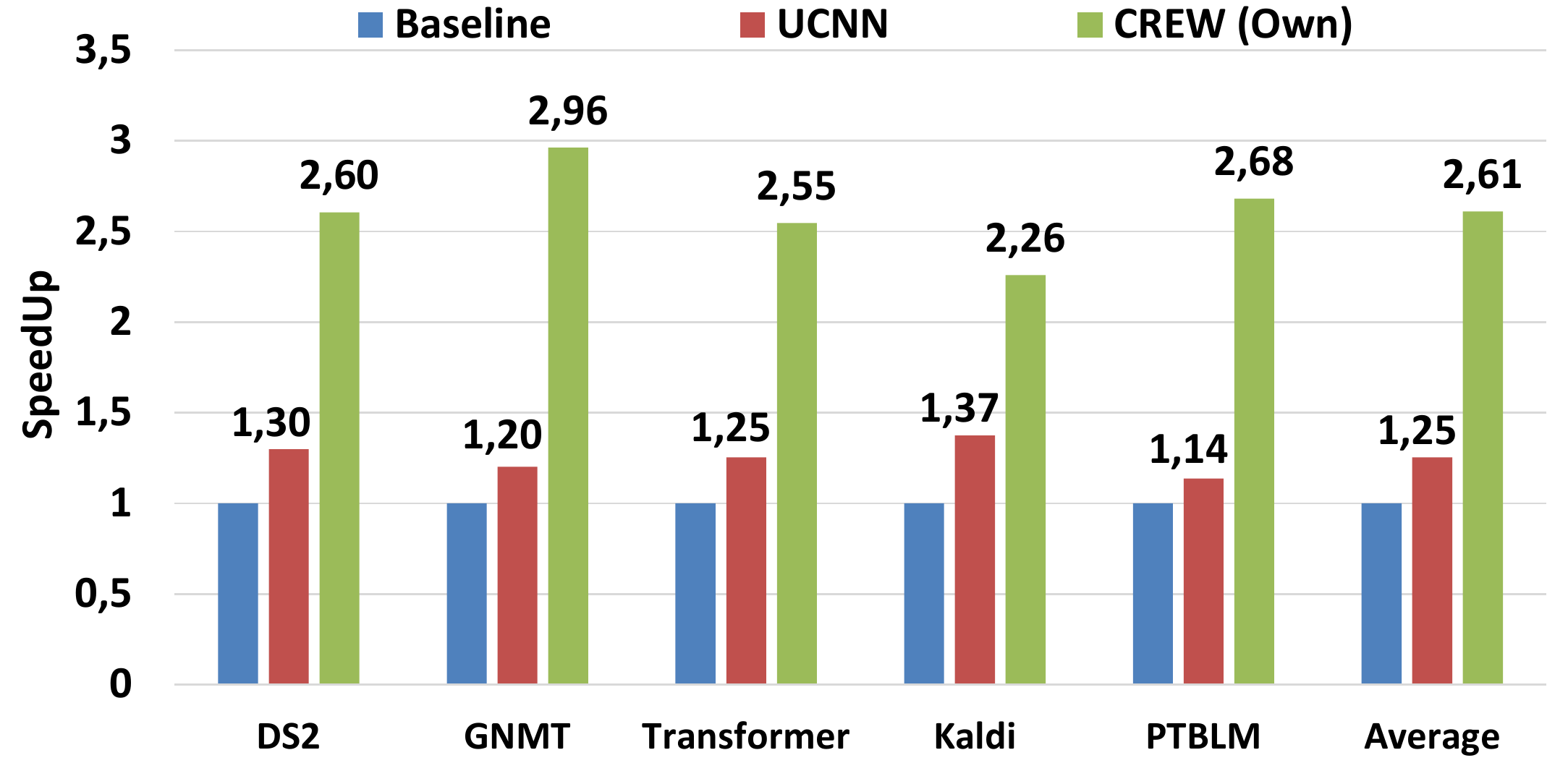}
\vskip -0.05in
\caption{Speedups achieved by CREW and UCNN for each DNN. Baseline configuration is the TPU-like DNN accelerator without any computation reuse mechanism.}
\vskip -0.20in
\label{f:speedup}
\end{figure}

Figure~\ref{f:speedup} shows the speedups achieved by CREW and UCNN over the baseline DNN accelerator. CREW provides consistent speedups for the five DNNs that range from $2.26x$ (\textit{Kaldi}) to $2.96x$ (\textit{GNMT}), achieving an average performance improvement of $2.61x$. The reduction in execution time is due to reusing previously computed partial products. The number of multiplications is dramatically reduced since only the unique weights are multiplied by their corresponding inputs. Furthermore, the overhead to access the previously memoized partial products is small since the indexes are smaller than the original weights. In addition, the partial sums are computed concurrently while loading the next blocks of indexes from memory, efficiently exploiting all the resources of the accelerator. Note that we use the partial product memoization scheme presented in Section~\ref{s:partial_product_mem}, so the speedups reported in Figure~\ref{f:speedup} are achieved without any accuracy loss. Impact on accuracy loss and performance of partial product approximation are presented in Section~\ref{s:crew_app_eval}.

We achieve more than $2x$ performance speedup compared to the factorization scheme of UCNN, since CREW is much more effective for FC layer inference. Although UCNN supports FC layers, our results show that it achieves an average speedup of $25\%$ compared to the baseline. The multiplications are reduced in a similar percentage compared to CREW. However, the overheads to index the inputs of each factorization group in an FC layer are quite high, hindering the benefits of the computation reuse scheme.

\begin{figure}[t!]
\centering
\includegraphics[width=0.75\columnwidth]{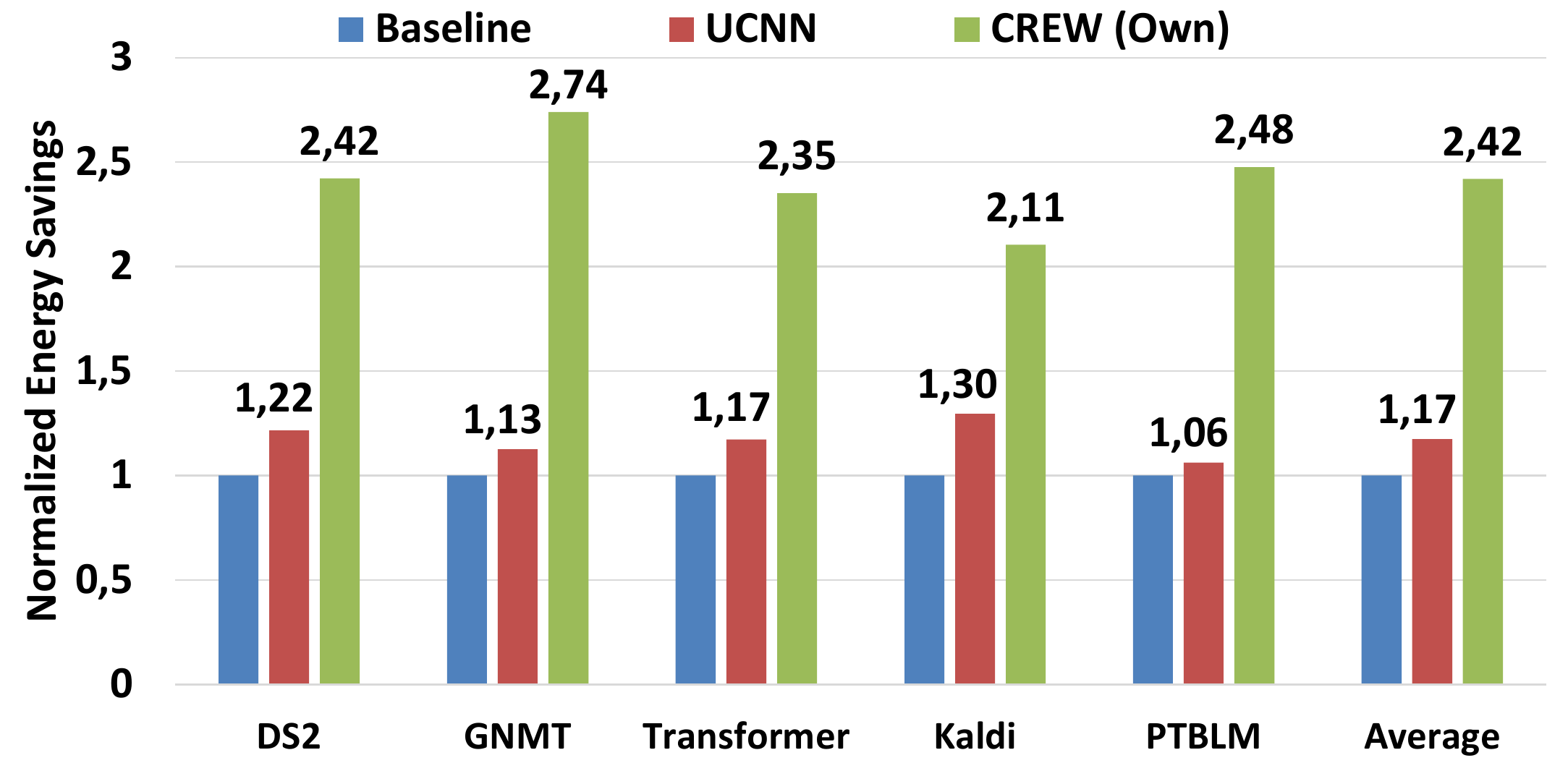}
\vskip -0.05in
\caption{Normalized energy savings for each DNN. Baseline configuration is the TPU-like DNN accelerator without the computation reuse technique.}
\vskip -0.20in
\label{f:energy_savings}
\end{figure}

Figure~\ref{f:energy_savings} reports normalized energy savings. On average, our scheme reduces the energy consumption of the accelerator by $2.42x$. The energy savings are well correlated with the reduction of the model size due to the low number of unique weights and the small size of the indexes compared to the original weights. These energy savings are due to two main reasons. First, dynamic energy is reduced due to the savings in computations and memory accesses. Second, the performance improvements shown in Figure~\ref{f:speedup} provide a reduction in static energy. Again, we reduce energy by more than $2x$ on average compared to both the baseline and UCNN.

\subsection{CREW-PPA Evaluation}\label{s:crew_app_eval}
As described in Section~\ref{s:partial_product_approx}, we propose to extend CREW with partial product approximation (CREW-PPA) to further reduce multiplications and the storage for the indexes. With this technique, some unique weights are discarded and replaced by the most similar remaining unique weights. Note that all the processing for this optimization is performed offline and it does not require any additional hardware support, so it can be implemented directly on top of CREW. This optimization effectively reduces the amount of computations and memory bandwidth usage, but it has to be carefully applied to guarantee a negligible impact on DNN accuracy.

\begin{figure}[t!]
\centering
\includegraphics[width=0.75\columnwidth]{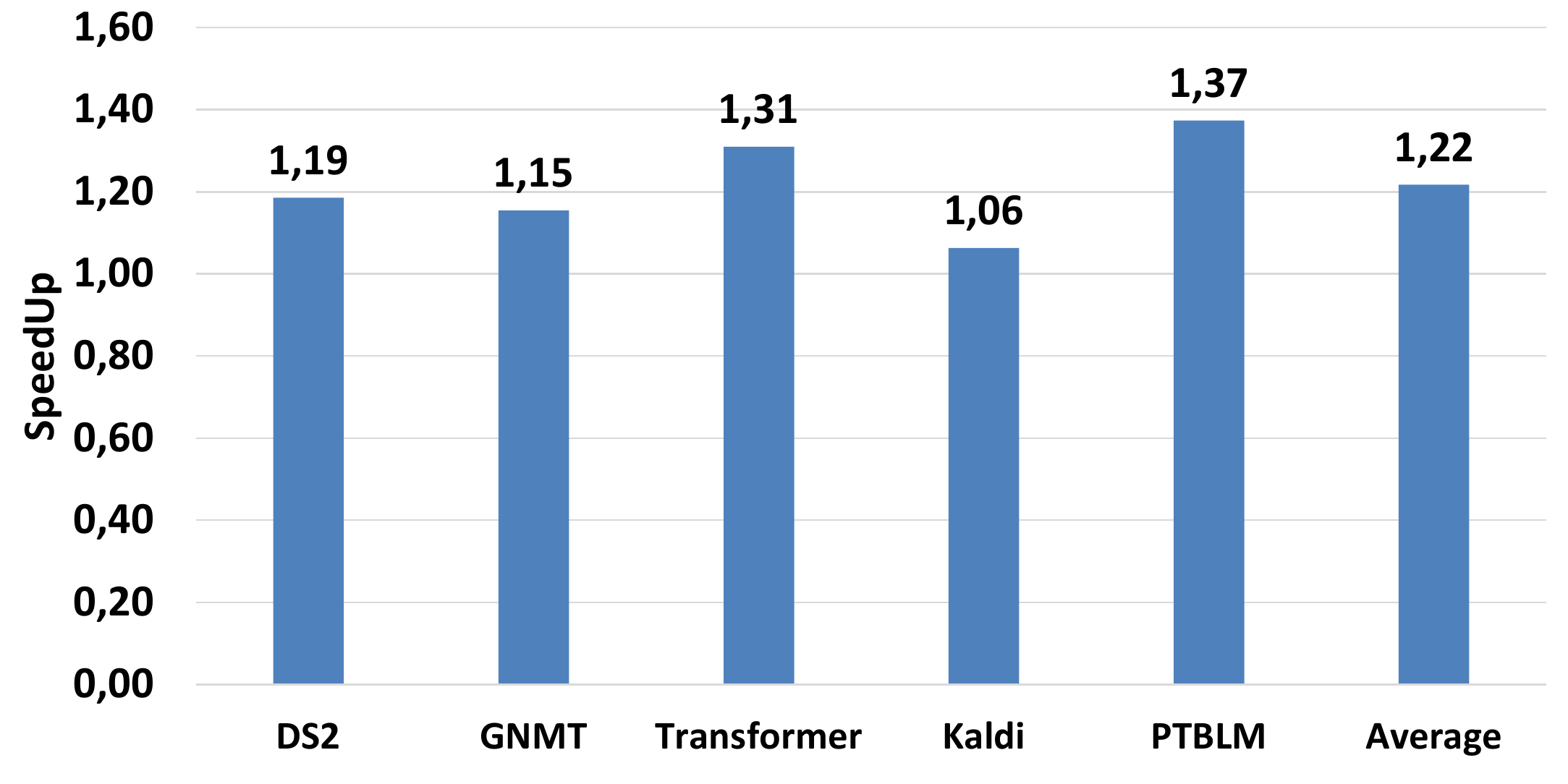}
\vskip -0.05in
\caption{Speedup of CREW-PPA with less than 1\% accuracy loss. Baseline configuration is the CREW accelerator.}
\vskip -0.20in
\label{f:speedup_approx}
\end{figure}

\begin{figure}[t!]
\centering
\includegraphics[width=0.80\columnwidth]{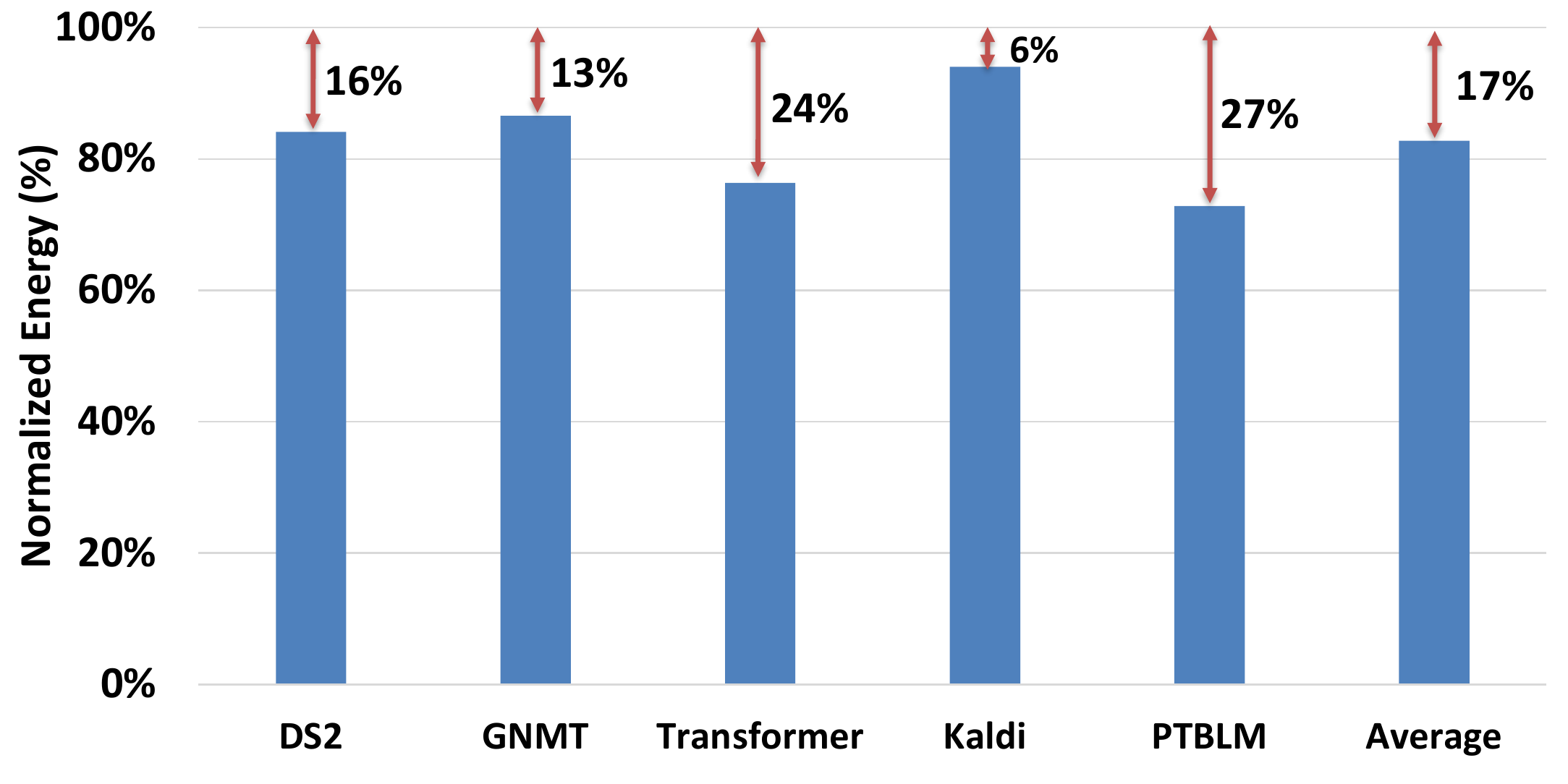}
\vskip -0.05in
\caption{Normalized energy of CREW-PPA with less than 1\% accuracy loss. Baseline configuration is the CREW accelerator.}
\vskip -0.25in
\label{f:energy_savings_approx}
\end{figure}

Figure~\ref{f:speedup_approx} and Figure~\ref{f:energy_savings_approx} show the speedup and normalized energy obtained of the PPA optimization when the maximum accuracy loss is set to 1\%. On average, it achieves more than $1.2x$ speedup with an energy reduction of 17\%, on top of the CREW accelerator. Note that this optimization can be customized to the user requirements in terms of accuracy versus performance and energy consumption.

\subsection{Overheads}\label{s:ov_eval}
CREW requires extra storage in the PEs of the accelerator as shown in Table~\ref{t:hw_params}. These buffers together with the index decoders are small compared to the global SRAM, so they represent a small portion of the area and energy of the accelerator. CREW has a 9\% increase in area, compared to the baseline TPU-like accelerator. On average, the energy consumed by the additional hardware of CREW represents less than 4\% of the total energy. In comparison, UCNN's extra storage results in an increase in area of 4\% and less than 3\% of energy consumption. We believe CREW's area and energy overheads are acceptable considering the large performance and energy improvements as reported in section~\ref{s:crew_eval}.

\section{Related Work}\label{s:related_work}

\textbf{DNN Optimizations}. Popular optimizations for reducing memory footprint and/or numerical precision of DNNs include clustering~\cite{K-Means} and quantization~\cite{original_qt}. TTQ~\cite{TTQ}, DoReFa~\cite{dorefa} and LQNets~\cite{lqnets} achieve important reductions in the numerical precision (i.e. $<8b$) with small impact in accuracy loss. However, clustering and quantization techniques do not reduce the amount of computations, but storage requirements and/or the cost of the computations. CREW supports these optimizations and efficiently exploits the weight repetition originated from them to improve performance and energy efficiency.

\textbf{Pruning and Sparse Accelerators}. Pruning~\cite{NearZeroPruning, Scalpel} reduces the model size and the number of computations by removing connections/nodes depending on the weights' values. The pruned model may loss accuracy but tends to regain it after retraining. On the other hand, the pruned model becomes sparse, requiring specialized hardware to be efficiently executed. Multiple sparse DNN accelerators~\cite{EIE, SCNN, CambriconX} have been recently proposed. Some sparse accelerators, such as EIE~\cite{EIE}, utilize clustering to further reduce weight storage, but do no explore ways to reuse computations as we propose in CREW. Note that CREW does not behave as a sparse accelerator since all weights are still processed in the form of small indexes and memoized products. Pruning is orthogonal to this work and could be applied to reduce the number of unique weights, further increasing the benefits of CREW.

\textbf{Computation Reuse}. Several recent studies observed the increase in repeated weights and/or inputs and proposed algorithms to reuse computations. UCNN~\cite{UCNN} and \cite{WeightSharingMAC} perform factorization of repeated weights in CNNs to reduce computations. However, as discussed in this work, FC layers have different requirements, and the proposed mechanisms do not fully exploit the benefits of computation reuse. Reuse of partial products with repeated weights has been previously proposed in \cite{CORN}, where the output neurons of a given FC layer are clustered depending on their weights similarity, and the partial products may be reused inside a given cluster of neurons. Note that CREW does not limit the reuse on a given cluster of neurons but can reuse partial products on any output neuron of an FC layer. The work in \cite{ReusePIM} proposes RAPIDNN, an In-Memory Processing (PIM) accelerator inspired by memoization and computation reuse techniques. RAPIDNN uses clustering to statically generate a subset of the most representative inputs and weights, and memoizes all the possible partial product combinations. During inference, the inputs and weights are encoded by accessing lookup tables to select which partial products are accumulated for each output. Note that the partial products selected are always an approximation, which may impact the accuracy of the network. Compared to our work, we do not loss any accuracy by performing linear quantization, and we require fewer accesses to lookup tables since the integer accumulation with our dataflow is efficiently performed in a systolic array architecture. On the other hand, a different approach is to exploit the repetition or similarity of input activations. Deep Reuse~\cite{DeepReuse} and \cite{ApproxReuse} detect similarities between vectors of inputs in a given CNN layer to approximate computations, while a recent proposal~\cite{InputSimilarity} exploits the similarity between consecutive frames of audio/video to reuse computations of consecutive executions of a given layer. These proposals do not exploit repeated weights and are orthogonal to our work.

\section{Conclusions}\label{s:conclusions}
In this paper, we show that modern DNNs exhibit a high degree of weight repetition in FC layers, resulting in a low number of unique weights per input neuron. Then, we propose CREW, a new accelerator that exploits weight repetition to reduce multiplications and memory accesses. The proposed reuse scheme only performs the multiplications of the inputs by their respective unique weights, avoiding many redundant computations and their associated memory accesses. The final output accumulations are performed by indexing a buffer that stores the results of the partial products. Besides, the indices required to access the buffer of partial products are small, since their size depend on the number of unique weights, thus the total network model size is significantly reduced. CREW also includes an enhanced weight stationary dataflow that processes blocks of indices and partial products. We show that CREW requires minor hardware changes over a state-of-the-art accelerator, mainly additional memory storage for saving the partial products and the indices. Our experimental results show that, on average, CREW provides $2.42x$ energy savings and $2.61x$ speedup, while it only requires a minor increase in the area of the accelerator (less than 9\%). We show that our scheme works for any DNN composed of FC layers such as MLPs and RNNs from different applications, including speech recognition and machine translation.


\section*{Acknowledgment}
This work has been supported by the CoCoUnit ERC Advanced Grant of the EU’s Horizon 2020 program (grant No 833057), the Spanish State Research Agency (MCIN/AEI) under grant PID2020-113172RB-I00, the ICREA Academia program, and the Spanish Ministry of Education under grant FPU15/02294.

\bibliographystyle{IEEEtran}
\bibliography{refs}

\end{document}